\documentclass[journal]{IEEEtran}

\usepackage{bm}
\usepackage{amsfonts}
\usepackage{amsmath}
\usepackage{amssymb}
\usepackage{lipsum}

\usepackage{multirow}
\usepackage{booktabs}

\usepackage[noadjust]{cite}

\usepackage{graphicx}
\usepackage{tikz}
\usepackage{pgfplots}
\pgfplotsset{compat=1.15}
\graphicspath{{./figure/}}

\makeatletter
\let\MYcaption\@makecaption
\makeatother

\usepackage[font=footnotesize]{subcaption}

\makeatletter
\let\@makecaption\MYcaption
\makeatother

\usepackage[hidelinks]{hyperref}
\usepackage{array}

\usepackage[linesnumbered, ruled, vlined]{algorithm2e}

\SetCommentSty{mycommfont}

\usepackage{algorithmic,algorithm2e,float}
		
\DeclareMathOperator*{\diag}{diag}

\newcommand{\R}{\mathbb{R}}

\newcommand{\X}{\mathfrak{X}}	 
\newcommand{\Z}{\mathfrak{Z}}	 

\newcommand{\bO}{\mathcal{O}}
\newcommand{\G}{\mathcal{G}}
\newcommand{\V}{\mathcal{V}}
\newcommand{\E}{\mathcal{E}}

\newcommand{\N}{\mathcal{N}}

\newcommand{\T}{\mathcal{T}}

\title{Graph Neural Networks Based Detection of Stealth False Data Injection Attacks in Smart Grids }

\date{\today}

\begin{document}
	\author{
		Osman~Boyaci$^{\ast}$,
		Amarachi~Umunnakwe$^{\ast}$,
		Abhijeet~Sahu$^{\ast}$,
		Mohammad~Rasoul~Narimani$^{\dagger}$,
		Muhammad~Ismail$^{\ddagger}$,
		Katherine~Davis$^{\ast}$, and
		Erchin~Serpedin$^{\ast}$
		\thanks{
			Manuscript received November 25, 2020; revised May 21, 2021; accepted August 19, 2021.
			\
			$\bm{\ast}$: Electrical and Computer Engineering, Texas A\&M University, College Station, TX, 77843 \texttt{\{osman.boyaci, amarachi, abhijeet\_ntpc, katedavis, eserpedin\}@tamu.edu};
			\
			$\bm{\dagger}$: College of Engineering, Arkansas State University, Jonesboro, AR, 72404 \texttt{mnarimani@astate.edu};
			\
			$\bm{\ddagger}$: Department of Computer Science, Tennessee Tech University, Cookeville, TN, 38505 \texttt{mismail@tntech.edu}.
			\
			This work was supported by NSF under Award Number 1808064. The data that support the findings of this study are available in \url{https://katedavis.engr.tamu.edu/projects/defenda/}

		}
	}

	\maketitle
	\begin{abstract}\label{abstract} 
	False data injection attacks (FDIAs) represent a major class of attacks that aim to break the integrity of measurements by injecting false data into the smart metering devices in power grids. To the best of authors' knowledge, no study has attempted to design a detector that automatically models the underlying graph topology and spatially correlated measurement data of the smart grids to better detect cyber attacks. The contributions of this paper to detect and mitigate  FDIAs are twofold.
	First, we present a generic, localized, and stealth (unobservable) attack generation methodology and publicly accessible datasets for researchers to develop and test their algorithms.
	Second, we propose a Graph Neural Network (GNN) based, scalable and real-time detector of FDIAs that efficiently combines model-driven and data-driven approaches by incorporating the inherent physical connections of modern AC power grids and exploiting the spatial correlations of the measurement. It is experimentally verified by comparing the proposed GNN based detector with the currently available FDIA detectors in the literature that  our algorithm outperforms the best available solutions by 3.14\%, 4.25\%, and 4.41\% in F1 score for standard IEEE testbeds with 14, 118, and 300 buses, respectively.
	\end{abstract}


	\section*{Nomenclature}
	\addcontentsline{toc}{section}{Nomenclature}
	\begin{IEEEdescription}[\IEEEusemathlabelsep\IEEEsetlabelwidth{$1234567890$}]
		\item[$P_i + jQ_i$] Complex power injection at bus $i$.
		\item[$P_{ij} + jQ_{ij}$] Complex power flow between bus $i$ and $j$.
		\item[$V_i, \theta_i$] Voltage magnitude and phase angle of bus $i$.
		\item[$\theta_{ij}$] $\theta_i - \theta_j$.
		\item[$G_{ij}+jB_{ij}$] $ij$th elements of bus admittance matrix.
		\item[$g_{ij}+jb_{ij}$] Series branch admittance between bus $i$ - $j$.
		\item[$g_{si}+jb_{si}$] Shunt branch admittance at bus $i$.
		\item[$\Omega_i$] Set of buses connected to bus $i$.
		\item[$\bm{z_o}, \bm{z_a} \in \R^{m}$] Original, attacked measurement vector.
		\item[$\bm{\hat{x}}, \bm{\check{x}} \in \R^{n}$] Original, attacked state vector.

		\item[$h(\bm{x})$] Nonlinear measurement function at $\bm{x}$.
		\item[$\bm{H} \in \R^{m \times n}$] Jacobian matrix.
		\item[$\bm{G} \in \R^{n \times n}$] Gain matrix.
		\item[$\bm{R}, \bm{S} \in \R^{m \times m}$] Error covariance, residual sensitivity matrix.
		\item[$\T$] Attacker's target area to perform FDIA.
		
	\end{IEEEdescription}

	\section{Introduction}\label{introduction} 

	As a highly complex cyber-physical system, a smart grid consists of a physical power system infrastructure and a cyber communication network. Physical measurement data are first acquired by the Remote Terminal Units (RTUs) or Phasor Measurement Units (PMUs) and are delivered to the Supervisory Control and Data Acquisition Systems (SCADAs). Then, the communication network transfers the measurement data to the application level where are processed and evaluated by the power applications  \cite{davis2012power}. Thus, reliability of power system depends on the security of the cyber-physical pipeline \cite{sridhar2011cyber}.
	
	Power system state estimation (PSSE) is a highly critical component of this pipeline since its outcome  is directly fed into numerous Energy Management System (EMS) blocks such as load and price forecasting, contingency and reliability analysis, and economic dispatch processes \cite{giannakis2013monitoring, abur2004power}. Thus, integrity and trustworthiness of the measurement data play a  critical role in ensuring proper operation of smart grids  \cite{he2016cyber}. By breaking this integrity, cyber-physical attacks target smart metering devices to harm the underlying physical systems. 
	
	False data injection attacks (FDIAs) represent a significant class of  cyber threats that modify   PSSE by maliciously altering the measurement data. In FDIAs, an attacker changes  sensor data in such a way that a valid and misleading operating point converge in PSSE and the attack becomes unobservable \cite{liang2016review}. Being unaware of the malicious data, the grid operator takes actions according to the false operating point of grid and consequently disrupts power system operation.   

	
	Traditional PSSE is performed using the weighted least squares estimation (WLSE) technique, and the presence of  bad data is detected by employing the largest normalized residual test (LNRT)  \cite{abur2004power}. Stealth (unobservable) FDIA can easily bypass the  bad data detection (BDD) systems. Therefore, FDIAs are one of the most critical attacks  for today's smart power systems.
	FDIAs in power grids were  first introduced a decade ago by \cite{liu2011false},  which  showed that an attacker with enough knowledge of the grid topology  can design an unobservable attack that satisfies the power flow equations and bypasses the BDD module.  Influential reference \cite{liu2011false} prompted an increased interest in detection of FDIAs \cite{chen2015detection, kosut2011malicious, huang2014real, drayer2019detection, s45, s60, s55, s65, s85, s92, s95, s99, s94, s97}.
	
	Most of the  works that deal with detection of FDIAs assume a linearized DC model  \cite{liu2011false, chen2015detection, kosut2011malicious, huang2014real, s45, s60, s65, s92, s94}. In the DC state estimation model, bus voltage magnitudes are assumed to be known as 1 p.u. and branch resistances and shunt elements are neglected. Hence, estimation of bus voltage angles is  reduced to linear matrix operations, and in general it helps to analyze the grid at some  extent. Although the linearized DC model is fast and simple, ignoring voltage magnitudes and reactive power components does not reflect the actual physical operation of the grid \cite{abur2004power}. Therefore, the  DC models can not validate that the FDIAs being tested are stealthy because PSSE and BDD tools employing AC power flow modeling can easily detect these  attacks without using extra detectors. 
	In addition, only a few works exploit  grid topology information into  their detection model \cite{drayer2018detection, drayer2019detection, ramakrishna2019detection} together with graph signal processing (GSP) techniques to detect FDIAs. Although innovative and powerful, these methods manually design  spectral filters, an operation which is not  scalable  since it requires manual and custom filter design steps. 
	Scalability is an essential feature that has to be considered when designing  detectors. Except a few highly scalable designs \cite{deng2015defending, liu2014detecting}, the majority of the proposed detectors for  FDIAs are designed for small scale systems such as IEEE 14 \cite{s45, s60, s65, s85} or IEEE 30 \cite{s95, s94}. Therefore, extensibility issues may arise when deploying  small-scall detectors at  large-scale networks.  
	Employing spatial-temporal correlations of the state variables and trust-based voting mechanisms, reference \cite{chen2015detection}  defines a consistency region and detection threshold to differentiate honest from malicious samples. Nevertheless, DC approximation and resolution of the time series data highly limit the applicability of the proposed design to  realistic large-scale power  grids. 

	Survey  \cite{musleh2019survey} classifies the FDIA detection algorithms  into two  categories:  model-based methods \cite{s45, s55, s60, s65, s85} and data-driven methods \cite{s92, s95, s99, s94, s97}. In general, model-based algorithms require first to build a  system model and estimate its parameters to detect FDIAs. Since there is no independent system to be trained, model-based methods do not need historical datasets;  nevertheless, threshold finding, detection delays and scalability aspects  restrict applicability of  model-based methods \cite{musleh2019survey}. On the contrary, data-driven models do not interfere with the system and its parameters, yet they necessitate historical data and a training process in order to reduce the detection time and increase scalability.

	Due to the superiority of machine learning (ML) methods along with the increasing volume of collected historical data samples, ML-based detectors have been proposed to identify FDIAs in smart grids. For example,  Decision Tree (DT) \cite{s97},   Support Vector Machine (SVM) \cite{s92},  \cite{s95} Multi Layer Perceptron (MLP) \cite{s95},  Recurrent Neural Network (RNN) \cite{s94},   Convolutional Neural Network (CNN) \cite{s99}  models were proposed to detect FDIAs. Despite their effectiveness, ML-based methods may overfit and fail to detect FDIAs especially in situations when the ML architecture does not capture the underlying physical system  generating the data \cite{musleh2019survey}. To illustrate, CNNs are well-suited to image and video processing since locality of  pixels is  well modeled by the sliding kernels.  Conversely, an RNN  architecture might be more applicable to recurrent relations such as sequence to sequence language modeling and machine translation applications \cite{deep_learning}.
	
	Undirected graphs can be used to capture the smart grid topology; buses and branches of the grid can be represented by nodes and edges of the undirected graph, respectively. The Graph Neural Network (GNN) architecture, in particular, immensely benefits from this architectural matching promise \cite{wang2020thirty, zhou2020graph}. Besides, the  prediction of the filter weights in GNNs instead of being performed manually  (e.g., \cite{drayer2018detection, drayer2019detection, ramakrishna2019detection}) can be executed automatically via GSP techniques which makes GNNs more attractive to  smart grid applications. For example, in \cite{owerko2020optimal}, GNNs are utilized for optimal power flow applications in power grids. Due to GNN's highly efficient modeling capability in non-Euclidean data structure, they are adopted in numerous areas such as social networks, physical systems, traffic networks, and molecule interaction networks \cite{zhou2020graph}. Despite their potential, to the best of our knowledge, no study has explored GNNs to detect FDIAs.
	
	In this paper, we propose a GNN-based stealth FDIA detection model for smart power grids. To fully model the underlying complex AC power system and dynamism of the measurements data, we decided to use a hybrid model; while system topology is integrated into our model by the help of GNN graph adjacency matrix, historical measurement data are modeled by the GNN spatial layers. These features enable  to take advantage of the benefits of both model-driven and data-driven approaches and hence better detect and mitigate FDIAs.

	The  contributions of this paper are summarized as follows:
		(1) We properly model the inherent cyber system: due to the topology and distribution of the smart measurement devices, meter readings are correlated in the measurement space of the smart grid; hence, ignoring the location of the meter data and assuming independent and identical distribution (iid) of meter readings may not be  accurate for a data-driven model. Therefore, we use GNN to match the cyber and physical layers of the grid. 
		(2) We design a stealth FDIA attack  methodology to test our detector: the main goal of any FDIA detector is to be able to detect stealth attacks since observable attacks can be easily detected by BDD systems. In other words, unproven random attacks do not require any extra detector other than traditional BDDs so the proposed detectors should be tested under stealth attacks to fully evaluate their performances. Therefore, we develop a Stochastic Gradient Descent (SGD) based stealth FDIA detection algorithm to exploit the possible weak points of the grid and  assess the performance in realistic conditions. It is experimentally  verified that the designed attacks can easily bypass classical BDD algorithms; however, they are detected by the proposed GNN detectors. 
		(3) We propose a scalable and real-time FDIA detector as an early warning/prediction system prior to the PSSE: since PSSE outcome is directly used by various EMS, the integrity of the measurements should be preserved. Thus, a detector system indicating the false data injection to the measurements prior to the PSSE is crucial. In addition, custom methods developed for small case systems may not be applicable to larger cases; therefore, detection models should be efficiently extensible to larger networks. Moreover, depending on the system scale and topology, detection delays can be very critical for power grids, therefore possible attacks should be detected as quickly as possible. Employing the standard test cases such as IEEE 14, 118, and 300 bus systems, it is demonstrated that the proposed method is linearly scalable both in parameter size and detection time.
	
	The remainder of this paper is divided into five sections. Section II is devoted to preliminaries such as power system state estimation, false data injection attacks and bad data detection mechanisms in smart grids. While Section III explains the proposed detection method and its mathematical modeling, Section IV describes the experimental results. 
	Finally, Section V concludes the paper. 
	
	\section{Power System Preliminaries} 
	
	
	\subsection{Power System State Estimation}

	PSSE module  aims to estimate the system state $\bm{x}$ ($V_i, \theta_i$ at each bus) in the steady state by using the complex power measurements $\bm{z}$ collected by noisy RTUs or PMUs via:
	\begin{equation} \label{eq:psse}
	\bm{\hat{x}} = \min_{\bm{x}} (\bm{z} - h(\bm{x}))^T \bm{R}^{-1} (\bm{z} - h(\bm{x})),
	\end{equation}	
	where  $\bm{R}$ denotes the error covariance matrix of measurements and $\bm{z}$ consists of active and reactive power injections at buses ($P_i, Q_i$) and active and reactive power flows on branches ($P_{ij}, Q_{ij}$). In polar form, these can be expressed as  \cite{abur2004power}:
	\begin{equation} \label{eq:pi}
	\begin{aligned}
	P_i &= \sum_{j \in \Omega_i} V_i V_j (G_{ij}\cos\theta_{ij} + B_{ij}\sin\theta_{ij}) = {P_G}_i - {P_L}_i \\
	Q_i &= \sum_{j \in \Omega_i} V_i V_j (G_{ij}\sin\theta_{ij} - B_{ij}\cos\theta_{ij}) = {Q_G}_i - {Q_L}_i \\
	P_{ij} &= V_i^2(g_{si} + g_{ij}) - V_iV_j(g_{ij}\cos\theta_{ij} + b_{ij}\sin\theta_{ij}) \\[0.3em]
	Q_{ij} &= -V_i^2(b_{si} + b_{ij}) - V_iV_j(g_{ij}\sin\theta_{ij} - b_{ij}\cos\theta_{ij}). \\
	\end{aligned}
	\end{equation}
	Since  (\ref{eq:pi}) are nonlinear and non-convex, (\ref{eq:psse}) is carried out via iterative weighted least squares estimation (WLSE)  \cite{handschin1975bad}.

	\subsection{False Data Injection Attacks}

	The goal of FDIA  is to find a new measurement vector $\bm{z_a}$ in the measurement space of the grid such  that  PSSE converges to another point in the state space of  variables. Formally,
	\begin{equation} \label{eq:attack}
	\bm{z_o} = h(\hat{\bm{x}}), \ \bm{z_a} = \bm{a} + \bm{z_o} = h(\check{\bm{x}}),
	\end{equation}
	where $\bm{a}$ represents the  attack vector, $\hat{\bm{x}}$ and $\check{\bm{x}}$ denote the estimated (original) state vector and false data injected state vector, and $\bm{z_o}$ and $\bm{z_a}$ stand for the  original and attacked measurements, respectively.
	
	\subsection{Bad Data Detection}
	Traditional power systems use the largest normalized residual test (LNRT) to detect bad samples using below eqs. \cite{abur2004power}:
	\begin{multline} \label{eq:bdd}
	\begin{split}
	& \bm{r} = \bm{z} - h(\bm{\hat{x}}), \  \bm{G} = \bm{H}^T \bm{R}^{-1} \bm{H},\\
	& \bm{S} = \bm{I} - \bm{H} (\bm{G}^{-1} \bm{H}^T \bm{R}^{-1} ), \ \bm{r}^N_i = \frac{|\bm{z}_i - h(\bm{\hat{x}})_i|}{\bm{R}_{ii} \bm{S}_{ii}}.
	\end{split}
	\end{multline}
	After estimating the current state vector using eq. (\ref{eq:psse}), residues $\bm{r}$ are calculated as the difference between observed ($\bm{z}$) and calculated ($h(\bm{\hat{x}})$) measurements. Then, using the Jacobian matrix $\bm{H}$ and the diagonal error covariance matrix of the measurements $\bm{R}$, gain matrix $\bm{G}$ is computed. Sensitivity of the residues for each measurement represented by the residual sensitivity matrix $\bm{S}$ are computed right after $\bm{G}$. Finally, residues are normalized by dividing each one of them with the product of corresponding diagonal elements of $\bm{R}$ and $\bm{S}$, and normalized residue vector $\bm{r}^N$ is obtained. Since $\bm{r}^N$ is assumed to have a standard normal distribution, a large $\bm{r}^N_i$ can be classified as bad data, if $\bm{r}^N_i$ exceeds a predetermined threshold $\tau_{bdd}$ specified by the grid operator according to the desired level of sensitivity \cite{abur2004power}. If an attacker wants to be considered stealthy, the maximum normalized residual value $\max(\bm{r}^N)$ should be less than the threshold $\tau_{bdd}$.

	\section{GNN Based Detection of FDIA}\label{idea} 
	
	\subsection{False data injection attack scenario}
\begin{figure}[h!]
		\centering
		\includegraphics[width=0.47\textwidth]{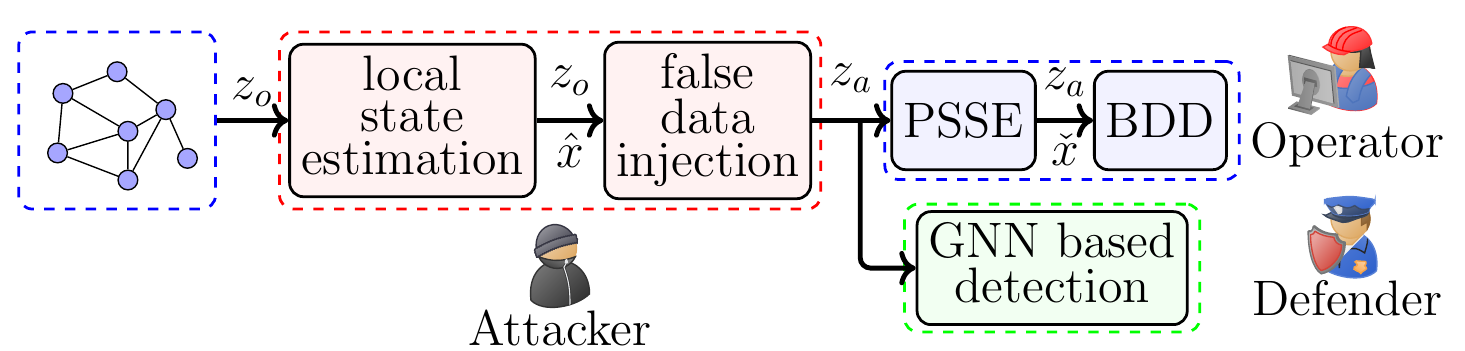}
		\caption{Architectural overview and signal flow graph of the proposed design. While blue boxes represent smart grids and their operations run by operator, red and green boxes denote functional blocks of attacker and defender, respectively. Note that operator gets attacked measurements $\bm{z_a}$ instead of original ones $\bm{z_o}$ due to the FDIA. Defender, on the contrary, tries to detect possible attacks by using $\bm{z_a}$.}
		\label{fig:overview}
	\end{figure}

	The main architecture and signal flow of the proposed design is illustrated in Fig. \ref{fig:overview}. 
	First, active and reactive power injections $P_i, Q_i$ at buses and active and reactive power flows $P_{ij}, Q_{ij}$ on branches are read by RTUs. Next, as a man in the middle, an attacker attempts to inject false data to the original measurements $z_o = [P_i, P_{ij}, Q_i, Q_{ij}]$ before the grid operator receives them. Then, using $\bm{z_a}$, the operator estimates the state variables and runs the BDD block to indicate a possible attack. In parallel, the defender runs the GNN-based detector 
	when it receives the measurements and hence predicts the probability of attack to warn the operator. In order not to raise suspicion from the operator, the attacker needs to design a stealth $\bm{z_a}$ that can bypass the BDD mechanism incorporated in eq. (\ref{eq:bdd}). At the same time, the attack strength should be strong enough to cause intended consequences or damages to the grid. In this regard, s/he initially estimates the state variables of grid in the target area $\T$, where security of the meters is compromised. Then, s/he searches a set of measurements $\bm{z_a}$ in the measurement space that serves the intended aim.
	
	As indicated by \cite{davis2012power, liu2011false}, FDIAs require that an adversary know the parameters and topology of the targeted portion of the system and is able to tamper the measurement data before the operator uses them in PSSE. Since accessing information and hardware 
	all over 
	the grid is neither easy nor realistic, we use a realistic `local' attack model to test our system. Due to the lack of open source, AC power flow based stealth FDIA generation algorithms to fully test the detection system, we propose a generic, localized AC stealth FDIA generation method using the stochastic gradient descent algorithm. 
	Herein scenario, the attacker focuses on a target area of the grid where the measurements s/he wants to inject the false data are located. To specify this area, it is assumed that s/he found an entry point $p$ in the cyber layer and can manipulate the measurements up to the  $r-$neighbor of $p$. Since generation buses and zero-injection buses would be too risky to change, s/he skips those buses even if they are in their active target region \cite{yang2013false, esmalifalak2011stealth, rahman2012false, yu2015blind}. Moreover, s/he avoids to attack the power flow measurements if this alternation leads to violate the KCL at the bus that the line is connected to \cite{hug2012vulnerability}.
	
\begin{figure}[h!]
		\centering
		\includegraphics[width=0.25\textwidth]{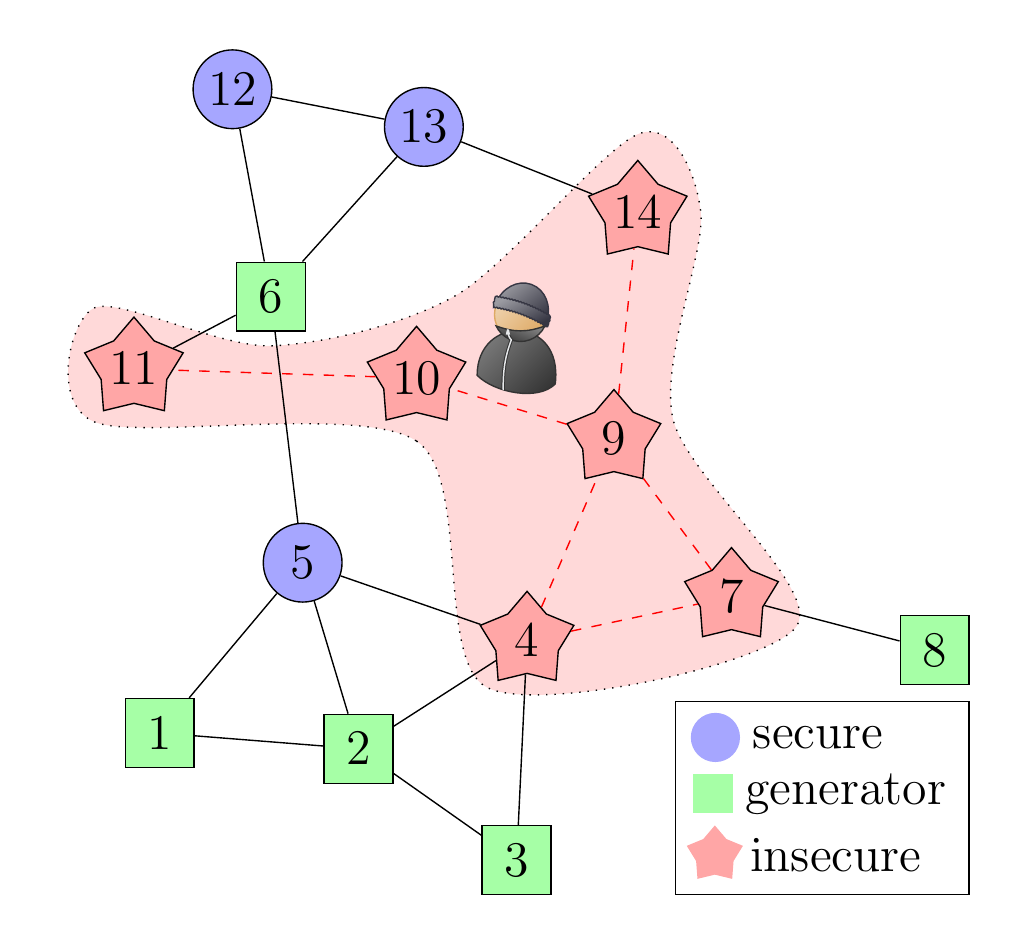}
		\caption{Visualization of an example IEEE 14 bus system where an attacker enters the system from bus 10 (entry point) and affects  the $1-$degree neighbors: bus 9 and 11, and $2-$degree neighbors: bus 4, 7, 14 depicted with red stars. Besides, it is assumed that s/he can change the power flows measurements depicted with red dashed lines in the target area $\T$ represented by the red surface. Note that since bus 6 is a generator node illustrated by a green square s/he skips it. Moreover, she avoids changing the line between bus 6 and 11 in order to not violate the KCL equation at 6.}
		\label{fig:attacker}
	\end{figure}

	An example IEEE 14 case system is demonstrated in Fig.~\ref{fig:attacker}, where an attacker's entry point $p$ is bus 10 and his radius $r$ is 2, so s/he can change the measurements of buses 10 (entry point), 9 and 11 ($1-$degree neighbors), and 4, 7, 14 ($2-$degree neighbors) designated with red stars. It is presumed that s/he can alter the measurements of the power flows in the active area represented by red dashed lines. Note that s/he skips bus 6 since it is a generator node designated by a green square. In addition, s/he also skips the line between 6 and 11 since, for this scenario, it is the only attackable meter connected to bus 6 and any change in this line violates the KCL equation at 6. All the other measurements outside the target region $\T$ represented by red surface are presumed to be still secure to the attacker.
	To find a stealth attack vector in $\T$, the attacker tries to minimize the objective function:
	\begin{equation} \label{eq:attack_opt}
	\begin{aligned}
	\min_{\check{\bm{x}}} \
	&  \lambda_{z} || h(\check{\bm{x}})_i - h(\hat{\bm{x}})_i ||_2 - 
	\lambda_{x} ||\check{\bm{x}}_{j} - \hat{\bm{x}}_{j}||,
	 \forall i \in \mathcal{T}_z,  \forall j \in \T_x\\
	\textrm{s.t.} \ 
	& h(\check{\bm{x}})_k = h(\hat{\bm{x}})_k, \ \check{\bm{x}}_{l} = \hat{\bm{x}}_{l},
	\ \forall k \not\in T_z, \ \forall l \not\in \T_x \\
	& \tau_{m}^{min} < ||\check{\bm{x}}|| < \tau_{m}^{max}, \ \tau_{a}^{min} < \angle{(\check{\bm{x}})} < \tau_{a}^{max} ,
	\end{aligned}
	\end{equation}	
	where $\hat{\bm{x}}$ denotes the honest state vector, $\check{\bm{x}}$ stands for  false data injected state vector, $\lambda_{z}$ and $\lambda_{x}$ are  weighting factors associated with loss terms, $\T_z$ and $\T_x$ denote the targeted measurements and state variables, $\tau_{m}^{min}$ and $\tau_{m}^{max}$ denote the minimum and maximum values of the magnitude of $\check{\bm{x}}$, and $\tau_{m}^{min}$ and $\tau_{m}^{max}$ represent minimum and maximum values of the angle of $\check{\bm{x}}$, respectively. In essence, s/he searches a vector $\check{\bm{x}}$ in the state space of the grid $\X$ by only targeting some $\bm{x} \in \T_x$ so that the corresponding measurements $\bm{z_a} = h(\check{\bm{x}})$ resemble  the original measurements $\bm{z_o}$ in the measurement space of the grid $\Z$ restricted by $\T_z$. Note that the objective function in  (\ref{eq:attack_opt}) consists of two competing losses. While the first part $||h(\check{\bm{x}})_i - h(\hat{\bm{x}})_i ||_2$ aims to minimize the measurement differences in $\T_z$,  the second part $||\check{\bm{x}}_{j} - \hat{\bm{x}}_{j}||$  maximizes the attack power injected into the state variables in $\T_x$. The trade-off between these objectives is directly related to detection risk and attack power since deviation from the original state variables increases the probability of being detected. Consequently, an attacker can increase the attack power at the expense of higher risk of being detected.
	
	The attacker aims to maximize the assault power by minimizing the detection risk. To do that, s/he first defines a free complex variable $\check{\bm{x}}_{j} \in \X$ in the vicinity of original estimated values by probing them with a small Gaussian noise. Then, by the help of SGD algorithm, s/he calculates the gradient of the state variables with respect to the joint loss defined in  (\ref{eq:attack_opt}) and updates them iteratively at each step until there is no improvement in the loss. Recall that s/he only updates a state variable if it is in the active insecure area. Eventually, s/he decides whether to inject this obtained false data to the related measurements in the cyber layer of the grid, according to the final loss value obtained during the iterations. In a sense, this individual latent vector search can be interpreted as `training' in the machine learning terminology \cite{glo}; however, it is very specific to the corresponding time slot and should be repeated for each case in order to minimize the detection risks. Note that this generic algorithm can be tailored according to the modeled electric grid and capabilities of the attacker.
		
	\subsection{Graph Neural Network Modeling of Smart Grids}	
	Smart power grids can be modeled by a connected, undirected, weighted graph $\G = (\V, \E, \bm{W})$ that consists of a finite set of vertices $\V$ with $|\V| = n$, a finite set of edges $\E$ and a weighted adjacency matrix $\bm{W} \in \R^{n \times n}$ \cite{ortega2018graph}. Buses are represented by vertices $\V$, branches and transformers are represented by edges $\E$ and line admittances are represented by $\bm{W}$ in this mapping. If the buses $i$ and $j$ are connected, the corresponding weight of the edge $e = (i, j)$ connecting vertices $i$ and $j$ is assigned to $W_{ij}$. A signal or a function $f:\V \rightarrow \R $ in $\G$ can be represented by a vector $\bm{f} \in \R^n$, where $i$th component of the vector $\bm{f}$ corresponds to scalar value at the vertex $i \in \V$.

	A fundamental operator defined in spectral graph theory \cite{ortega2018graph} is the graph Laplacian operator $L \in \R^{n \times n}$. Its normalized definition is represented as
 	$L  = I_n - D^{-1/2} W D^{-1/2}$ 
	where $I_n$ is the identity matrix, and $D \in \R^{n \times n}$ is the diagonal degree matrix with $D_{ii} = \sum_{j} W_{ij}$. Since $L$ is a real symmetric positive semi-definite matrix, all eigenvalues $\lambda_i$ of it are real valued and non-negative, and it has a complete set of orthonormal eigenvectors $\bm{u_i}$ \cite{ortega2018graph}. Thus, $L$ can be diagonalized as $L = U \Lambda U^T$
	where $U = [\bm{u}_0, \bm{u}_1, \ldots, \bm{u}_{n-1}] \in \R^{n \times n}$ represent the $n$ orthonormal eigenvectors, and $\Lambda = \diag([\lambda_0, \lambda_1, \ldots, \lambda_{n-1}]) \in \R^{n \times n}$ denotes the diagonal matrix of $n$ eigenvalues  $0=\lambda_0 < \lambda_1 < \ldots < \lambda_{n-1} < 2$ due to the normalization \cite{ortega2018graph}. In fact, vectors $\bm{u}_i$ form the graph Fourier basis and $\lambda_i$ values represent frequencies in the graph spectral domain \cite{ortega2018graph}.
	
	The Fourier Transform and its inverse can be defined in the graph spectral domain analogously to the classical Fourier Transform. Namely, the Graph Fourier Transform (GFT) and Inverse Graph Fourier Transformation (IGFT) are defined as
	$ \tilde{s} = U^T s$ and $s = U \tilde{s}$
	where $s$ and $\tilde{s}$ denote vertex and spectral domain signals, respectively.

	Unlike classical signal processing, a meaningful translation operator does not exist in the vertex domain \cite{gcnn}. Therefore, to apply a convolution operation to graph signals, they are first transformed into the spectral domain using GFT, then convolved (Hadamard product) in the spectral domain and finally the result transformed back to the vertex domain using IGFT \cite{gcnn}. Formally, $x \ast_\G y = U ((U^T x) \odot (U^T y))$.
	
	Similarly, a graph signal $x \in \R^{n}$ is filtered by a kernel $g_\theta$:
	\begin{equation} \label{eq:filter}
	\begin{aligned}
		y = g_\theta \ast_\G x = g_\theta(U \Lambda U^T) x = U g_\theta(\Lambda) U^T x \in \R^{n},
	\end{aligned}
	\end{equation}
	 where $g_\theta(\Lambda) = \diag(\theta)$ is a non-parametric kernel, and $\theta \in \R^n$ is a vector of Fourier coefficients \cite{ortega2018graph}. To put it differently, $g_\theta$ filters the signal $x$ in the spectral domain by multiplying its spectral components with the free $\theta$ coefficients in a similar way with the classical signal processing in the Fourier domain. Eventually, the filtered signal is transformed back to the vertex domain by IGFT \cite{ortega2018graph}. Nevertheless, those non-parametric filters are not  spatially  localized and hence computational complexity of eq. (\ref{eq:filter}) is $\bO(n^2)$ due to the matrix multiplication with $U$. To thwart this problem, \cite{gcnn} proposed to parameterize $g_\theta(L)$ as a Cheybyshev polynomial function which can be computed recursively from $L$. 
	 
	 The $K$ order Chebyshev polynomial of the first kind $T_k(x)$ is computed recursively as follows  \cite{mason2002chebyshev}:	
	\begin{equation} \label{eq:cheb_recur}
		T_k(x) = 2x T_{k-1}(x) - T_{k-2}(x),
	\end{equation}
	where $T_0(x)=1$ and $T_1(x)=x$.
	Therefore, a filter $g_\theta$ can be approximated by 
	Chebyshev polynomials, $T_k$, up to order $K-1$ and a signal $x$ can be filtered by  $g_\theta$:
	\begin{equation} \label{eq:graph_conv_cheby}
	\begin{aligned}
		y = g_\theta \ast_\G x = g_\theta(L) x = \sum_{k=0}^{K-1} \theta_k T_k(\tilde{L}) x,
	\end{aligned}
	\end{equation}
	where the parameter $\theta \in \R^K$ is a vector of Chebyshev coefficients, and $T_k(\tilde{L}) \in \R^{n \times n}$ is the Chebyshev polynomial of order $k$ evaluated at the scaled Laplacian $\tilde{L}$ given by $\tilde{L} = 2 L / \lambda_{max} - I_n$.
	Finally, filtered signal $y$ can be calculated by the help of  (\ref{eq:cheb_recur}) and (\ref{eq:graph_conv_cheby}) as:
	\begin{equation} \label{eq:final_res}
		y = \sum_{k=0}^{K-1} \theta_k \bar{x}_k,
	\end{equation}
	where $\bar{x}_0=x$, $\bar{x}_1=\tilde{L} x $, and $\bar{x}_k$ is computed recursively:
	\begin{equation} \label{eq:final_recur}
	\bar{x}_k = 2\tilde{L}\bar{x}_{k-1} - \bar{x}_{k-2}.
	\end{equation}
	
	Note that convolution in eq. (\ref{eq:final_res}) is $K$-localized, and its computational complexity is reduced to $\bO(K|\E|)$. Thus, it can efficiently be utilized in the intermediate layers of the GNN to model the non-Euclidean measurement data of power grids. For detailed analysis, please refer to \cite{gcnn, ortega2018graph}. 
	
	\subsection{Detection of Attacks Using Graph Neural Network}

\begin{figure}[h]
		\centering
		\includegraphics[width=0.47\textwidth]{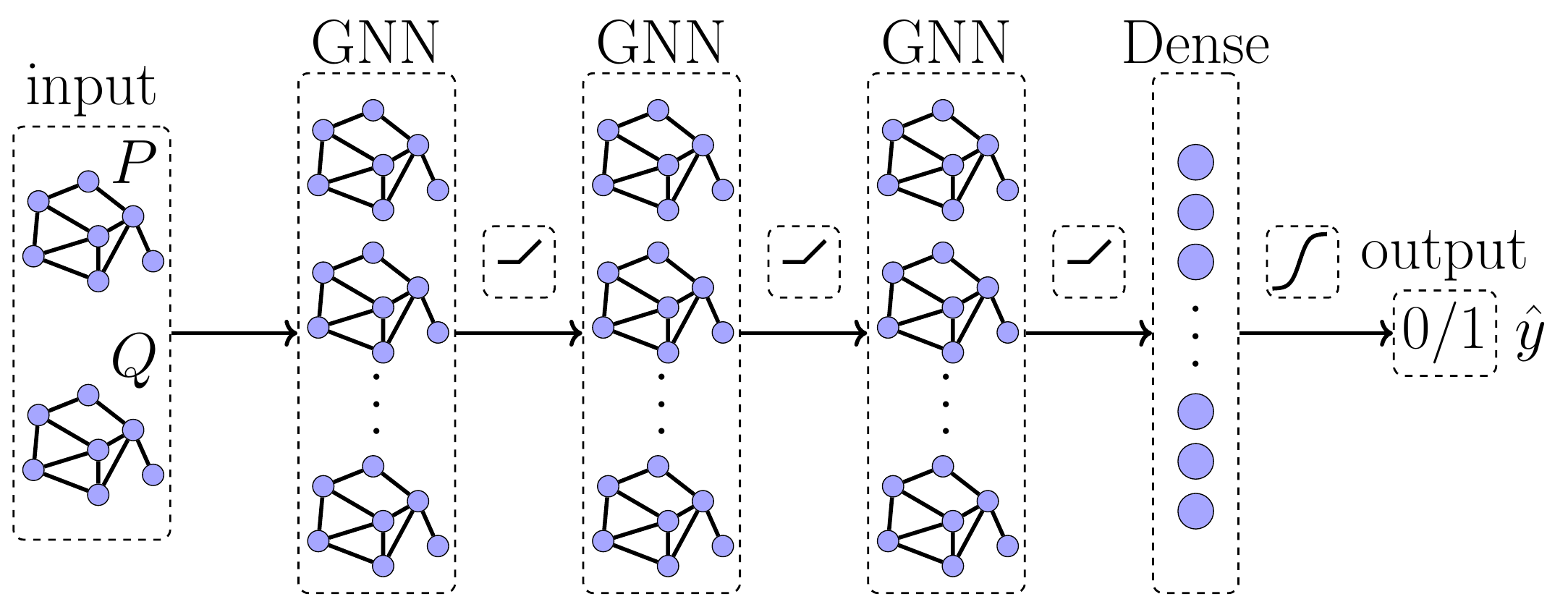}
		\caption{Architecture of the proposed GNN based detector.}
		\label{fig:arhitecture}
	\end{figure}

	The architecture of the proposed GNN-based detector is depicted in Fig.~\ref{fig:arhitecture}. It contains one input layer to represent bus power injection measurements, $L$ hidden Chebyshev graph convolution layers to extract spatial features and one output dense layer to predict the probability of the input sample being attacked. In this layered structure, $X^0$ denotes two channel input tensor $[P_i, Q_i] \in \R^{n \times 2}$, $X^l$ represents the output tensor of  hidden layer $l \in \R^{n \times c_l}$, $y \in \R$ designates the scalar output of the neural network, $1 \leq l \leq L$, and $c_l$ stands for the number of  channels in layer $l$. Particularly, a GNN hidden layer $l$ takes $X^{l-1} \in \R^{n \times c_{l-1}}$ as input and produces $X^{l} \in \R^{n \times c_{l}}$ as output. Different from the hidden graph layers, dense layer outputs $y$ in classical feed-forward neural networks  by feeding with the inputs $X^{L} \in \R^{n \times c_{L}}$.
	In this multi-layer architecture, each Chebyshev layer $l$ for $1 \leq l \leq L$ transforms its input $X^{l-1}$ by first applying graph convolution operation using eqs. (\ref{eq:final_res}) and (\ref{eq:final_recur}), then adding a bias term and finally employing a nonlinear rectified linear unit function (ReLU) defined as $\textrm{ReLU}(x) = \max(0, x)$ to generate $X^l$. Namely, 
	\begin{equation} \label{eq:layer}
		X^l = \textrm{ReLU}(\theta^l \ast_\G X^{l-1} + b^l),
	\end{equation}
	where $\theta^l \in \R^{K \times c_{l-1} \times c_{l}}$ denotes free Chebyshev coefficients and $b^l \in \R^{c_l}$ represents bias term of the layer $l$ . Recall that each Chebyshev layer gets extra scaled Laplacian $\tilde{L}$ values. In a similar fashion, output of the dense layer is computed by $y = \sigma(W^L X^L + b^L)$, where $W^L \in \R^{n \times c_L} $ denotes the weights of each feature, $b^L \in \R$ represents the bias term and $\sigma$ designates the nonlinear sigmoid operation: $\sigma(x) = 1/(1+e^{-x})$.   
	
	\section{Experimental Results}\label{details} 
	\subsection{Data Generation}
\begin{figure}[]
		\centering
		\begin{subfigure}{.24\textwidth}
			\centering
			\includegraphics[width=.99\linewidth]{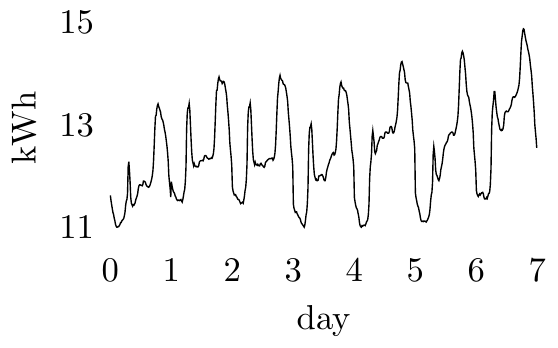}
			\caption{Load profile of south-central Texas}
			\label{fig:scaler_x}
		\end{subfigure}
		\begin{subfigure}{.24\textwidth}
			\centering
			\includegraphics[width=.99\linewidth]{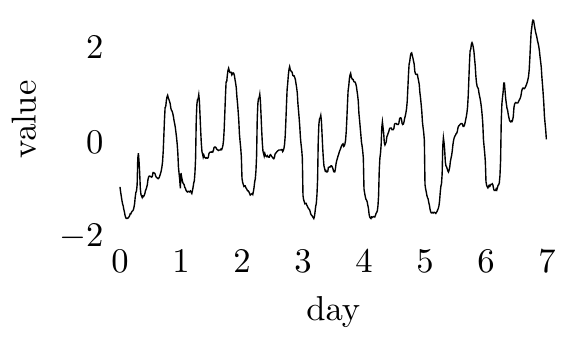}
			\caption{Scaler $\bm{S}$}
			\label{fig:scaler_k}
		\end{subfigure}
		\begin{subfigure}{.24\textwidth}
			\centering
			\includegraphics[width=.99\linewidth]{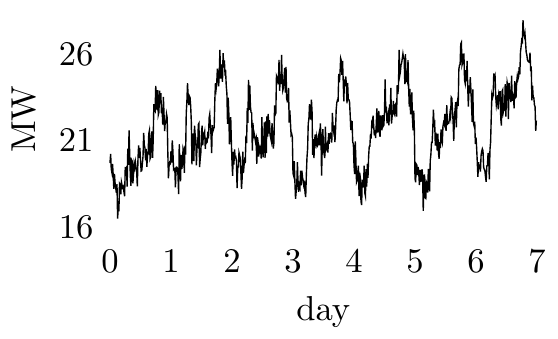}
			\caption{Load of bus 2 in IEEE-14}
			\label{fig:scaler_l}
		\end{subfigure}
		\begin{subfigure}{.24\textwidth}
			\centering
			\includegraphics[width=.99\linewidth]{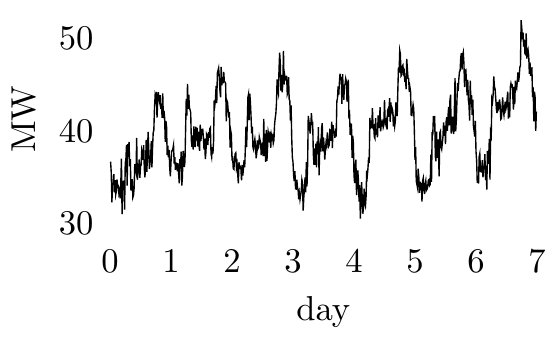}
			\caption{Generation of bus 2 in IEEE-14}
			\label{fig:scaler_g}
		\end{subfigure}
		\caption{An example scaling process for bus 2 in the IEEE-14 bus test system. First, load profile of south-central Texas region (Fig.~\ref{fig:scaler_x}) is normalized and the scalar $\bm{S}$ is obtained (Fig.~\ref{fig:scaler_k}). Then, load (Fig.~\ref{fig:scaler_l}) and generation (Fig.~\ref{fig:scaler_g}) values of buses are multiplied with a scalar value sampled from the distribution $\N(1+ k \times \bm{S_t},\, \sigma_s)$ having $1+ k \times \bm{S_t}$ mean and $\sigma_s$ standard deviation at time $t$. For this example, the static values of load and generation at bus 2 in the IEEE-14 bus system are 21.7 MW and 40 MW, respectively. Please note that while $S$ has relatively smooth transition between the time-steps, load and generation values have some spikes due to deviation of multiplier around $\bm{S_t}$ which increases the variety of samples in the data-set.}
		\label{fig:scaler}
	\end{figure}

	Generating reliable data is the first step in building a successful defense mechanism since all the future blocks depend on it. Since it is not possible to find publicly available power grid data due to privacy issues, synthetic data are generated using Pandapower \cite{pandapower} for several test cases including IEEE 14, 118, and 300. Data generation steps are summarized in Algorithm~\ref{alg:data}. To make the data as realistic as possible, we first downloaded ERCOT's 15 minutes interval backcasted actual load profiles \cite{ercot}.
	\begin{algorithm}[h] 
 		\algsetup{linenosize=\scriptsize}
 		\small
		\caption{Data generation}
		\label{alg:data}
		\SetKwInOut{Input}{Input}
		\SetKwInOut{Output}{Output}
		\SetKwFunction{FMain}{Main}
		\SetKwFunction{FGen}{Generate}
		\DontPrintSemicolon
		
		\Input{normalized scaler $\bm{S}$ \tcp*{$\mu = 0, \ \sigma = 1$}}
		\Output{$\bm{Z_n}, \bm{X_n}$ for each test system $n$}
		$N \gets [14, \ 118, \ 300]$ \tcp*{IEEE bus systems}
		$T \gets [1$ \KwTo $9600]$   \tcp*{timestep index}
		$k, \ \sigma_s \gets 0.1, \ 0.03$ \tcp*{scaling coefficients}
		$\sigma_n \gets 0.01$ \tcp*{noise coefficient}
		
		\SetKwProg{Fn}{Function}{:}{}
		\Fn{\FGen{$sg, \ t$}}{
			\ForEach{$bus \in sg.gen bus \cup sg.load bus$}{%
				$bus.scale \gets \N(1+ k \times \bm{S_t},\, \sigma_s)$\;
			}
			$\bm{z_o} = sg.PF()$ \tcp*{run AC power flow}
			$\bm{z_o} \gets \N(\bm{z_o}, \ \bm{z_o} \times \sigma_n) $,		\tcp*{1\% additive noise}
			$\hat{\bm{x}} \gets sg.PSSE(\bm{z_o})$ \tcp*{estimate state}
			\KwRet $\bm{z_o}, \ \hat{\bm{x}}$ \;
		}
		
		\SetKwProg{Fn}{Function}{:}{\KwRet}
		\Fn{\FMain}{
			\ForEach{$n \in N$}{
				$\bm{Z_n},  \ \bm{X_n} \gets [ \ ],  \ [ \ ]$ \tcp*{empty vectors}
				$sg \gets $ SG($n$) \tcp*{smart grid obj.}
				\ForEach{$t \in T$}{%
					$\bm{z}, \ \bm{x} \gets$ Generate($sg, \ t$)\;
					$\bm{Z_n[t]}, \ \bm{X_n[t]} \gets z, \ x$ \tcp*{append}
				}
				$\bm{Z_n}.save(), \ \bm{X_n}.save()$
			}
		}
	\end{algorithm}
	Next, we arbitrarily selected the `BUSHILF\_SCENT' profile which corresponds to south-central  Texas having a high load factor.
	Then, we normalized the time series data to  zero mean and unit variance `scaler' vector $\bm{S}$ so that it can be easily adapted to each test system. Having obtained $\bm{S}$, we run the Main function of the Algorithm~\ref{alg:data} where a smart grid object $sg$ is created for each test system having $n$ bus and Generate function is called for each timesteps $t$.
	In Generate function, the scaling parameters of  generator and load buses are assigned to a sample drawn from a normal distribution with $1 + 0.1 \times \bm{S_t}$ mean and $0.03^2$ variance, where $\bm{S_t}$ denotes the value of $\bm{S}$ at time-step $t$.
	Due to the properties of  normal distribution, the scaling operation provides practically more than $\pm20\%$ dynamic range on average with respect to the static case.
	We limit the scaling range between 0.7 and 1.3 for the convergence of power flow solutions.
	As a next step, AC power flow solutions are calculated, and the measurements considered to have 1\% noise are read.
	Finally, PSSE is conducted, and estimated state variables are returned along with original meter values to the Main function.
	In Fig.~\ref{fig:scaler}, the scaling process formulated with line 7 in Algorithm \ref{alg:data} is demonstrated for one week period ($7 \times 96$ samples).
	Please note that $\bm{S}$ depicted in Fig.~\ref{fig:scaler_k} is just a normalized version of the load profile given in Fig.~\ref{fig:scaler_x}.
	Next, the load and generation values of buses are multiplied with a value sampled from a distribution $\N(1+ k \times \bm{S_t},\, \sigma_s)$ which has $1+ k \times \bm{S_t}$ mean and $\sigma_s$ standard deviation at time $t$.
	Namely, they follow the patterns in $\bm{S}$ by deviating around their static values defined in their test systems.  

	\subsection{Attack Generation}
	After generating honest data samples, we focus on malicious data samples in this subsection, where the attack generation steps are summarized in Algorithm \ref{alg:attack}. The algorithm gets original measurements matrix $\bm{Z_n} \in \R^{T \times m} $ and estimated state variable matrix $\bm{X_n} \in \R^{T \times n}$ and produces their attacked version as well as corresponding sample vector $Y_n \in \R^T$, where 0 and 1 in $\bm{Y_n}$ represent honest and malicious samples, respectively. As can be seen from Algorithm \ref{alg:attack}, Main function simply creates the smart grid and attacker objects, fetches the current sample and calls Generate function for each system having $n$ buses at each time-step $t$.
	
	Generate function, in contrast, simulates a `smart' intruder capable of entering the cyber layer of the grid, computing an unobservable attack vector and deciding to insert the false data into the measurement devices according to the `quality' of the attack. In this regard, since it is not realistic to assume that an attacker can inject false data at every time step due to practical reasons, Generate function first models the attack frequency by a r.v. $f \sim \N(0, 1)$ where $f>\tau_{freq}$ means the attacker has successfully entered the system. To attack roughly 15\% of total time-steps on average, $\tau_{freq}$ is selected as 1. Second, it models the target area of the attacker $\T$ similar to the red area given with Fig.~\ref{fig:attacker} by help of a r.v. $p \sim \mathcal{U}(1,n)$ and a predefined attack radius $r$. To this end, it calls a breadth first search (BFS) method of the attacker object to model the target area defined by a set of measurements captured by the attacker denoted by $\T_z$ and a set of state variables $\T$ intended to inject the false data. In fact, all the measurements and state variables located up to $r$-distance neighbor of the bus $p$ are assumed to be in $\T_z$ and $\T_x$ except the generator buses and zero-injection buses. Then, it calls the attack method of the attacker to compute and insert  $z_a$ if the method returns  a $loss$ value smaller than  threshold $\tau_{loss}$.
	
	\begin{algorithm}[h] 
 		\algsetup{linenosize=\scriptsize}
 		\small 
		\caption{Attack generation}
		\label{alg:attack}
		\SetKwInOut{Input}{Input}
		\SetKwInOut{Output}{Output}
		\SetKwFunction{FMain}{Main}
		\SetKwFunction{FGen}{Generate}
		\SetKwFunction{FAtt}{attacker.attack}
		\DontPrintSemicolon
		
		\Input{$\bm{Z_n}, \ \bm{X_n}$ for each test system $n$}
		\Output{$\bm{Z_n}, \ \bm{X_n}, \ \bm{Y_n}$ for each test system $n$}
		$ N \gets [14, \ 118, \ 300]$ \tcp*{IEEE bus systems}
		$ T \gets [1$ \KwTo $9600]$   \tcp*{timestep index}
		$ \sigma_n \gets 0.005$ \tcp*{initial disturbance}
		$ \lambda_{z}, \ \lambda_{x} \gets 1, \ 1$ \tcp*{loss weights}
		$ \eta , \ E \gets 0.001, 1000$ \tcp*{learning rate and epochs}
		$\tau_{freq}, \ \tau_{loss} \gets 1, \ 0.1$ \tcp*{attackers thresholds}
		$R_{min} \gets \{14:2, \ 118:3, \ 300:6\}$ \tcp*{min radius}
		$R_{max} \gets \{14:3, \ 118:4, \ 300:8\}$ \tcp*{max radius}
		
		\SetKwProg{Fn}{Function}{:}{}
		\Fn{\FAtt{$\bm{z_o}, \ \hat{\bm{x}}, \ \T_z, \ \T_x$}}{
			$ \bm{trainable} \ \bm{V} : 0.9<\bm{V}<1.1$\;
			$ \bm{trainable} \ \bm{\theta} : -\pi<\bm{\theta}<+\pi$\;
			$ \bm{V}, \ \bm{\theta} \gets \text{abs} (\hat{\bm{x}}), \ \text{angle} (\hat{\bm{x}}) $ \;
			
			\ForEach{$j \in \T_x$}{
				$ \bm{V_j} \gets \bm{V_j} + \N(0, \ \sigma_n^2) $\;
				$ \bm{\theta_j} \gets \bm{\theta_j} + \N(0, \ \sigma_n^2)$\;	
			}
			
			\ForEach{$epoch \in E$}{
				$\check{\bm{x}} \gets \bm{V} e^{j \bm{\theta}} $  \tcp*{complex state vars.}
				$\bm{z_a} \gets h(\check{\bm{x}})$ \tcp*{real measurements}
				$L_z \gets \sum_{i} || {\bm{z_a}}_i - {\bm{z_o}}_{i} ||_2, \  \forall i \not\in \T_z$ \;
				$L_x \gets \sum_{j}  ||\check{\bm{x}}_{j} - \hat{\bm{x}}_{j}||, \ \forall j \not\in \T_x$ \;
				$L \gets \lambda_{z} L_z - \lambda_{x} L_x $\;
				\ForEach{$j \in \T_x$}{
					$ \bm{V_j} \gets \bm{V_j} - \eta \frac{\partial L}{\partial \bm{V_j}} $\;
					$ \bm{\theta_j} \gets \bm{\theta_j} - \eta \frac{\partial L}{\partial \bm{\theta_j}}$\;	
				}
			}
			$\check{\bm{x}} \gets \bm{V} e^{j \bm{\theta}} $  \tcp*{complex state vars.}
			$\bm{z_a} \gets h(\check{\bm{x}})$ \tcp*{real measurements}
			\KwRet $\bm{z_a}, \ L$ \;
		}
		
		\SetKwProg{Fn}{Function}{:}{}
		\Fn{\FGen{$attacker, \ \bm{z_o}, \ \hat{\bm{x}} $}}{
			
			$y, \ \bm{z} \gets 0, \ \bm{z_o}$		\tcp*{no attack yet}
			$f \sim \N(0, \, 1)$ 				\tcp*{attack frequency}
			\If{$ f > \tau_{freq}$}{%
				$ p \sim \mathcal{U}(1,n)$ \tcp*{entry point}
				$ r \gets \mathcal{U}(R_{min}[n], \ R_{max}[n])$			    \tcp*{attack radius}
				\tcc{determine attack surface by BFS}
				$\T_z, \ \T_x \gets attacker.BFS(p, r)$
				$\bm{z_a}, \ loss \gets attacker.attack(\bm{z_o}, \ \hat{\bm{x}}, \ \T_z, \ \T_x)$\;
				\If{$loss < \tau_{loss}$}{
					$y, \ \bm{z} \gets 1, \ z_a$		\tcp*{attack injected}
				}
			}
			$\check{\bm{x}} \gets sg.PSSE(\bm{z})$ \;
			\KwRet $\bm{z}, \ \check{\bm{x}}, \ y$ \;
		}
		
		\SetKwProg{Fn}{Function}{:}{\KwRet}
		\Fn{\FMain}{
			\ForEach{$n \in N$}{
				$\bm{Y_n} \gets [ \ ]$ \tcp*{empty label vector}
				$sg \gets $ SG($n$) \tcp*{smart grid obj.}
				$attacker \gets Attacker(n)$ \tcp*{attacker obj.}
				\ForEach{$t \in T$}{%
					$\bm{Z_n[t]}, \ \bm{X_n[t]}, \ \bm{Y_n[t]} \gets$ Generate($attacker, \ \bm{Z_n[t]}, \ \bm{X_n[t]}$)\;
				}
				$X_n.save(), \ Z_n.save(), \ Y_n.save()$
			}
		}
	\end{algorithm}

\begin{figure}[]
		\centering
		\newcommand{\wdt}{0.155} 
		\begin{subfigure}{\wdt\textwidth}
			\centering
			\includegraphics[width=.99\linewidth]{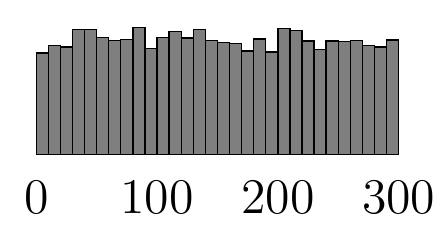}
			\caption{Entry point}
			\label{fig:x}
		\end{subfigure}
		\begin{subfigure}{\wdt\textwidth}
			\centering
			\includegraphics[width=.99\linewidth]{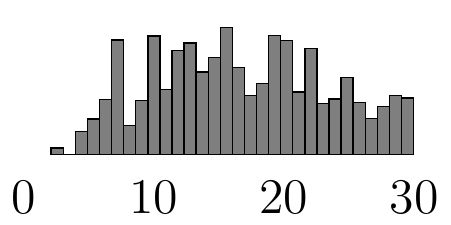}
			\caption{Hacked meter $\%$}
			\label{fig:r}
		\end{subfigure}
		\begin{subfigure}{\wdt\textwidth}
			\centering
			\includegraphics[width=.99\linewidth]{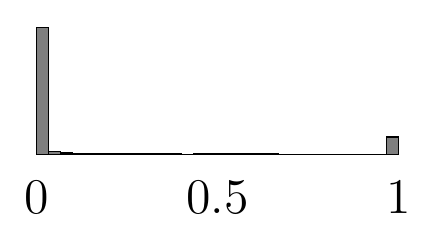}
			\caption{Minimization loss}
			\label{fig:loss}
		\end{subfigure}
		\begin{subfigure}{\wdt\textwidth}
			\centering
			\includegraphics[width=.99\linewidth]{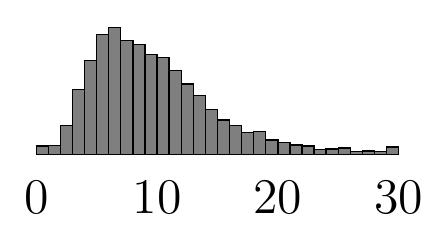}
			\caption{$\max\Delta|z|$[mw,mvar]}
			\label{fig:z}
		\end{subfigure}
		\begin{subfigure}{\wdt\textwidth}
			\centering
			\includegraphics[width=.99\linewidth]{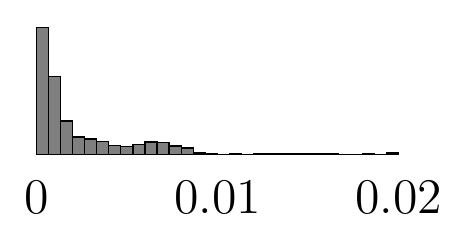}
			\caption{$\max \Delta |V|$ [p.u.]}
			\label{fig:vm}
		\end{subfigure}
		\begin{subfigure}{\wdt\textwidth}
			\centering
			\includegraphics[width=.99\linewidth]{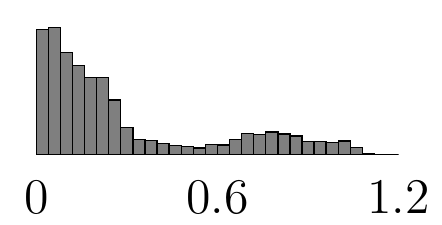}
			\caption{$\max \Delta |\theta|$ [degree]}
			\label{fig:va}
		\end{subfigure}
		\caption{Distributions of attacker's entry point (\ref{fig:x}), ratio of seized meters in percentage (\ref{fig:r}), minimization loss values by solving eq.~(\ref{eq:attack_opt}) (\ref{fig:loss}), maximum absolute difference of attack to measurements (Fig.~\ref{fig:z}) and maximum absolute difference of state variables due to attacks in terms of voltage magnitude (\ref{fig:vm}) and angle (\ref{fig:va}) obtained by Algorithm~\ref{alg:attack}. Roughly speaking, the attacker initiates the assault arbitrarily and uniformly from any node (Fig.~\ref{fig:x}) by capturing up to 30\%  of the available meters (Fig.~\ref{fig:r}) and succeeding 83.4\% of the attempts (Fig.~\ref{fig:loss}). Adding maximum 30 MW or 30 MVAR of attack to measurement devices (Fig.~\ref{fig:z})  creates maximum  2\% deviation in magnitudes (Fig.~\ref{fig:vm}) and maximum of 1.2 degree in angles (Fig.~\ref{fig:va}) of  state variables. Attack frequency and attack power might be strengthened by increasing  $\tau_{freq}$ and $\tau_{loss}$ parameters in Algorithm \ref{alg:attack} at  risk of high detection by  operator.}
		\label{fig:attack_hist}
	\end{figure}

	The attacker's assault method solves the nonlinear and non-convex minimization    (\ref{eq:attack_opt}) in the Tensorflow \cite{tensorflow} library. As a first step, it defines a free trainable vector tuple to represent the new complex state variables $\check{\bm{x}}$ which  constitutes the `fake' operating point at the end of attack:  voltage magnitude $\bm{V}$ is limited to $0.9<\bm{V}<1.1$ p.u.  and voltage angle $\bm{\theta}$ is limited to  $-\pi<\bm{\theta}<\pi$. Next, it initializes the $j$th elements of this tuple in the vicinity of their original variables by adding a small Gaussian white noise $\N(0, \ \sigma_n^2)$ if $j \in \T_x$ to ignite the optimization. This small proximity could play a vital role because SGD may fail to reduce the objective function if the initial point is not balanced \cite{glo}.  A  $\bm{\check{\bm{x}}}$ too close to  $\bm{\hat{\bm{x}}}$ might result to no update at all in optimization variables $\bm{V}$ and $\bm{\theta}$, whereas a $\bm{\check{\bm{x}}}$ too distant to  $\bm{\hat{\bm{x}}}$ might get stuck in a secluded region of $\X$ and produce a highly suspicious $\bm{z_a}$. Thus, $\sigma_n = 0.005$ is found to be accurate according to the minimization loss. Then, for each epoch, it obtains $\bm{z_a}$ using $h(\bm{x})$ and consequently calculates loss term  $L_z$ as a root mean squared error between $\bm{z_a}$ and $\bm{z_o}$, and $L_x$ as a mean absolute error between $\check{\bm{x}}$ and $\hat{\bm{x}}$. Eventually, it calculates gradients of total loss $L = \lambda_{z} L_z - \lambda_{x} L_x$ with respect to optimization variables $\bm{V_j}$ and $\bm{\theta_j} \in \T_x$ and updates corresponding terms in the reverse  direction of gradients by scaling the gradients with learning rate $\eta$ before starting the next epoch. Lastly, it returns $\bm{z_a}$ and final loss $L$  to  Generate function and halts.
	Distributions of some important values of  IEEE 300 test system are given in Fig.~\ref{fig:attack_hist} after running Algorithm~\ref{alg:attack}.
	
	\subsection{Attack Detection}
	In order to immediately predict the attack probability in our models instead of waiting for  PSSE result, we only use measurement values in our detectors. Moreover, since $P_i+jQ_i = \sum_{k \in \Omega_i} P_{ik} +jQ_{ik}$, node values can represent branch values as summation in their corresponding $\Omega_i$ and the proposed GNN-based detector accepts features in its nodes, we decide to use only $P_i$ and $Q_i$ as input to our models. PSSE and BDD modules, on the contrary, continue to receive every available measurement to operate as depicted in Fig.~\ref{fig:overview}.
	
	Having decided to input features $[P_i, Q_i]_n \in \R^{9600 \times n \times 2}$ and output labels $ \bm{Y_n} \in \R^{9600}$ for $n \in {\{14, 118, 300\}}$ bus test systems where 0 denotes honest and 1 denotes malicious samples of $\bm{Y_n}$, we partition the first 60\% of the samples for training the proposed detectors, the next 20\% for validating and tuning the hyper-parameter of the models, and the last 20\% for evaluating the performances of the detectors. Then, we standardize each split separately, with a zero mean and a standard deviation of one, to have a faster and more stable learning process \cite{machine_learning}.
	
	As a next step, we implement the GNN-based FDIA detector having a multi-layer Chebyshev graph convolution layer in its hidden layer and one dense layer on top of that as depicted in Fig. \ref{fig:arhitecture}. We add a bias term and ReLU activation functions between graph convolutional layers and sigmoid activation functions at the last dense layer to increase the detector's nonlinear modeling ability \cite{machine_learning}.
	As for weighted adjacency matrix $\bm{W}$, we use the magnitude of complex sparse Ybus matrix of the corresponding grid, which models the relation between nodes, determine the graph Laplacian $L$ and scale it to  obtain $\tilde{L}$.

	All free unknown parameters defined in the model are computed by a supervised training using cross-entropy loss:
	\begin{equation} \label{eqn:train}
	L(\hat{y}, W_{\theta}) = \frac{-1}{N} \sum_{n=1}^{N}	y_i \log(\hat{y}_i) + (1-y_i)\log(1-\hat{y}_i),
	\end{equation}
	over the training set where $N$ denotes the number of samples in the training set, $W_{\theta}$ represents all trainable parameters $\theta_l$ and $b_l$ for $1 \leq l \leq L$ along with $W^L$ and $b^L$ in the model, and $y_i$ and $\hat{y}_i$ stand for true and predicted class probability for sample $i$, respectively. Training samples are fed into the model as mini batches having 64 samples with 128 maximum number of epochs in addition to the early stopping where 16 epochs are tolerated without any improvement in the cross entropy loss of validation set. All the implementation was carried out in Python 3.8 using Pandapower~\cite{pandapower},  Sklearn~\cite{sklearn}, and Tensorflow~\cite{tensorflow} libraries on Intel i9-8950 HK CPU \@ 2.90GHz with NVIDIA GeForce RTX 2070 GPU.
	
	To evaluate the performance of proposed model in the  binary classification task, we use  true positive rate or detection rate (DR)  $DR=TP/(TP + FN)$ as  probability of attack detection,  false positive rate or false alarm rate (FA)  $FA=FP/(FP + TN)$ as  probability of falsely alarming the system even though there is no attack, and the F-measure or F1 score $F_1=2*TP/(2TP + FP + FN)$ as the harmonic mean of the precision and sensitivity of  classifier \cite{machine_learning}, where TP, FP, TN, and FN stand for true positives, false positives, true negatives and false negatives, respectively. 
	
	\begin{figure}[h!]
		\centering
		\newcommand{\wdt}{0.15} 
		\begin{subfigure}{\wdt\textwidth}
			\centering
			\includegraphics[width=.99\linewidth]{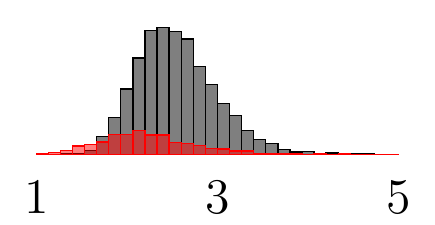}
			\caption{$\bm{r}^N$ for IEEE 14}
			\label{fig:014_bdd}
		\end{subfigure}
		\begin{subfigure}{\wdt\textwidth}
			\centering
			\includegraphics[width=.99\linewidth]{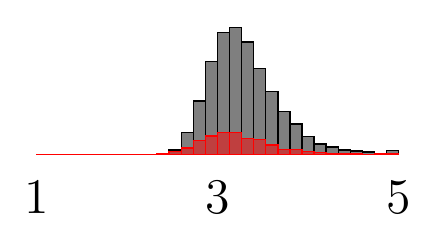}
			\caption{$\bm{r}^N$ for IEEE 118}
			\label{fig:014_gnn}
		\end{subfigure}
		\begin{subfigure}{\wdt\textwidth}
			\centering
			\includegraphics[width=.99\linewidth]{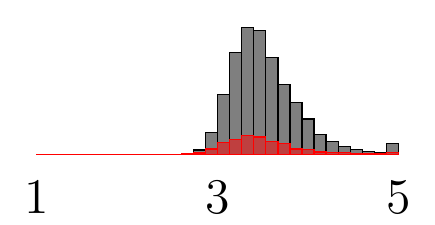}
			\caption{$\bm{r}^N$ for IEEE 300}
			\label{fig:118_bdd}
		\end{subfigure}
		\begin{subfigure}{\wdt\textwidth}
			\centering
			\includegraphics[width=.99\linewidth]{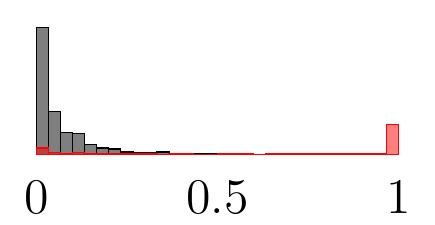}
			\caption{$\hat{y}$ for IEEE 14}
			\label{fig:118_gnn}
		\end{subfigure}
		\begin{subfigure}{\wdt\textwidth}
			\centering
			\includegraphics[width=.99\linewidth]{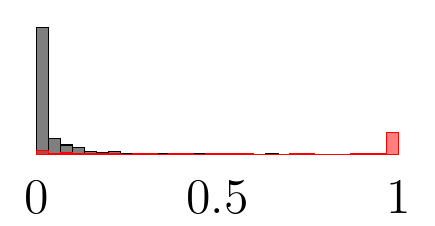}
			\caption{$\hat{y}$ for IEEE 118}
			\label{fig:300_bdd}
		\end{subfigure}
		\begin{subfigure}{\wdt\textwidth}
			\centering
			\includegraphics[width=.99\linewidth]{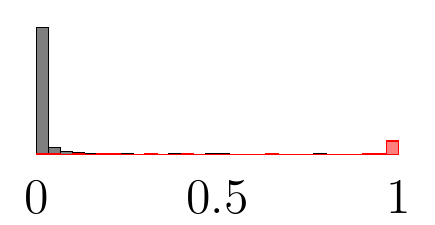}
			\caption{$\hat{y}$ for IEEE 300}
			\label{fig:300_gnn}
		\end{subfigure}
		\caption{Distributions of normalized residues $\bm{r}^N$ and predicted class probabilities $\hat{y}$ for each IEEE test system having 14, 118, and 300 buses, respectively, computed on test dataset where black and red bars denote honest and malicious samples, respectively. GNN based detector transforms the class distributions  so that they can be easily separated. In traditional BDD, in contrast, it is not possible to isolate the bad samples due to stealthy FDIA.}
		\label{fig:class}
	\end{figure}
	
	Predicted class probabilities $\hat{y}$ obtained by GNN-based detector along with $\bm{r}^N$ values computed by the LNRT-based BDD system are given side-by-side for each test system in Fig.~\ref{fig:class}. Note that while it is almost impossible to separate honest and malicious samples by $\bm{r}^N$ due to intricate class distributions on the left side, the proposed GNN-based detector efficiently `filters' malicious samples in its hidden layers and provides easily separable $\hat{y}$ distributions. Please refer to Table~\ref{tab:results} for detailed classification results.
	
	\subsection{Model Scalability}
\begin{figure}[h!]
		\centering
		\includegraphics[width=.5\linewidth]{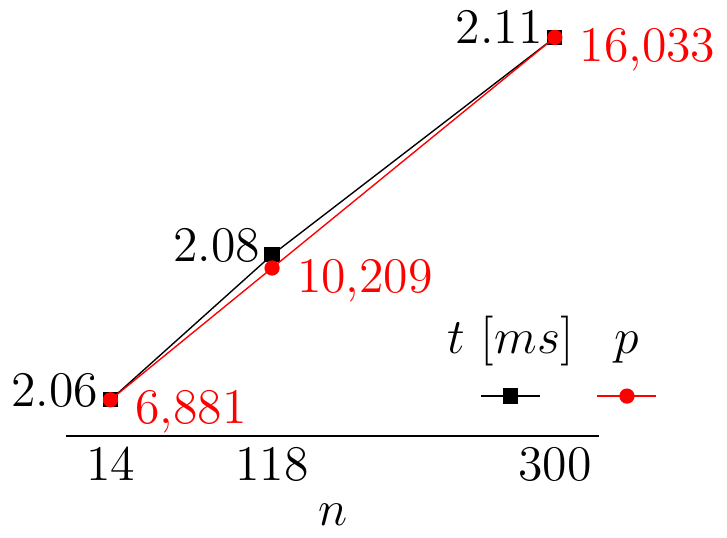}
		\caption{Linear scalability of the proposed models in terms of prediction time $t \ [ms]$ and total number of parameters $p$.}
		\label{fig:scalability}
	\end{figure}

	Model scalability in terms of total number of parameters and prediction time is examined herein subsection. We first assess the total number of free trainable parameters in the proposed models. While each $K$-localized Chebyshev layer $l$ having $c_l$ channels for $1 \leq l \leq L$ consists of $K \times c_{l-1} \times c_{l}$ Chebyshev coefficients and $c_l$ bias terms, the final dense layer assumes  $n \times c_L$ dense weights and a bias term. Thus, the total number of parameters in the model is given by:
	\begin{equation} \label{eq:param}
		K \sum_{l=1}^{l=L} ((c_{l-1}+1) \times c_{l}) + n \times c_L + 1.
	\end{equation}
	It can be seen from  (\ref{eq:param}) that except for  the last dense layer, the number of parameters in a GNN is free from bus size $n$ and it linearly and independently scales with the neighborhood order $K$, previous layer's filter size $c_{l-1}$ and its own filter size $c_l$. Second, we measure and save the prediction delays of each system. To fairly analyze how prediction time $t$ and total number of parameters $p$ change with the increasing bus size $n$, we fix the other variables at $K=3$, $L=3$, and $c_l=32$ for each layer $l$. Fig.~\ref{fig:scalability} demonstrates that models are linearly scalable in terms of $n$.

	\subsection{Visualization of how information spreads through layers}
\begin{figure}[h!]
		\centering
		\newcommand{\wdt}{0.24} 
		\begin{subfigure}{\wdt\textwidth}
			\centering
			\includegraphics[width=.99\linewidth]{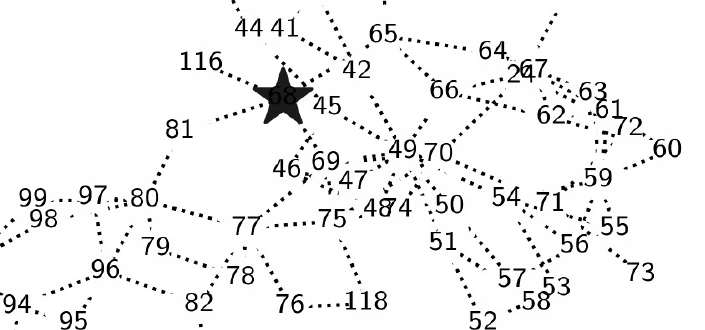}
			\caption{Input layer difference $\bm{\delta}^0$}
			\label{fig:filter0}
		\end{subfigure}
		\begin{subfigure}{\wdt\textwidth}
			\centering
			\includegraphics[width=.99\linewidth]{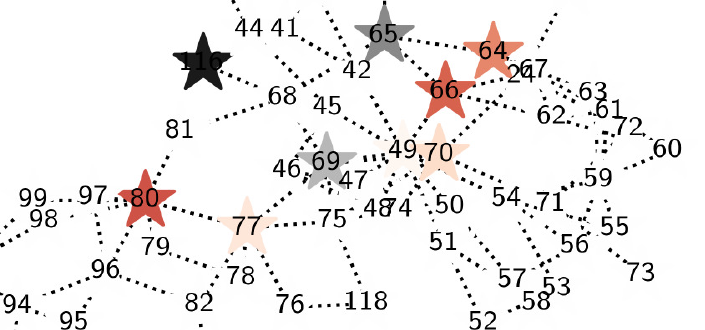}
			\caption{A layer 1 filter difference $\bm{\delta}^1$}
			\label{fig:filter1}
		\end{subfigure}
		\begin{subfigure}{\wdt\textwidth}
			\centering
			\includegraphics[width=.99\linewidth]{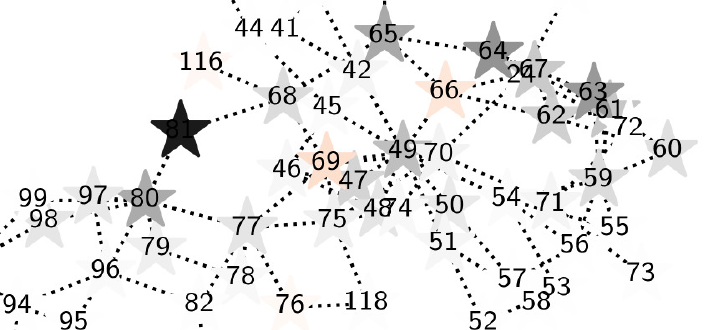}
			\caption{A layer 2 filter difference $\bm{\delta}^2$}
			\label{fig:filter2}
		\end{subfigure}
		\begin{subfigure}{\wdt\textwidth}
			\centering
			\includegraphics[width=.99\linewidth]{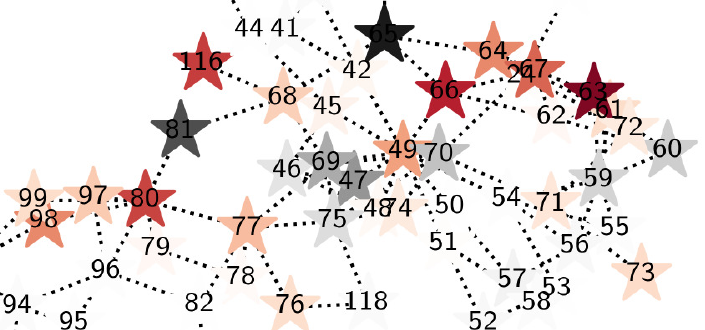}
			\caption{A layer 3 filter difference $\bm{\delta}^3$}
			\label{fig:filter3}
		\end{subfigure}
		\caption{Visualization of anomaly propagation through the layers of the proposed network. To better follow how the anomaly spreads, we plotted the signal differences at each layer. It can be clearly seen that each node sends information up to its $K$-neighbor in each filter and the message advances $K$ nodes through each layer. Note that $K$ is chosen 3 for this network in the training phase. For clarity, only affected area of the grid is depicted.}
		\label{fig:filter}
	\end{figure}

	In this subsection, to explain and visualize how the proposed GNN-based detector distinguishes a malicious sample from an honest one, we examine the output of the filters from each layer of a trained network. To this end, first we arbitrarily select a node from the center region of the grid, for instance, bus 68 of the IEEE 118 bus system. Second, we randomly choose an honest sample $\bm{s}$ from the training data set and create a malicious sample $\acute{\bm{s}}$ by adding a point-wise attack to the bus 68 with a magnitude one to easily follow the spreading of this anomalous information through the hidden layers of the network. To focus on the anomaly, we calculate the difference of the Chebyshev filter outputs $\bm{\delta}^l = \acute{\bm{s}}^l - \bm{s}^l$ at each layer $l$ for $0 \leq l \leq 4$ including the input layer $l=0$. Starting from $\bm{\delta}_0$, the example filter output differences from each of the Chebyshev layers are depicted in Fig.~\ref{fig:filter}. It can be clearly seen from Fig.~\ref{fig:filter} that each node transmits this anomalous data to its $K$-neighbors, where $K$ is chosen as 3 for this network, and the information advances in $K$-locality through each of the Chebyshev layers. The dense layer at the end of the model, in contrast, uses the anomalous features and decides its outputs by a sigmoid function. As expected, $\bm{s}^4 = 0$ and $\acute{\bm{s}}^4 = 1$ at the output of the model. In essence, $K$-localized Chebyshev filters of the proposed detector extract this spatial information through its GNN and dense layers to predict the probability of attacks for the input sample.  

	\subsection{Comparison with Other Methods}
	
	To compare our GNN-based models with the available detectors, we also implement Decision Tree  (DTC) \cite{s97}, Support Vector  (SVC) \cite{s92}, Multi Layer Perceptron  (MLP) \cite{s95}, Recurrent Neural Network  (RNN) \cite{s94}, and Convolutional Neural Network  (CNN) \cite{s99} based FDIA detectors. Since we do not have access to the data set of corresponding works, we train, validate and test these models similar to our proposed detector using our dataset. 
	
	DTC is a member of the non-parametric and supervised machine learning algorithms family aiming to create a multitude of decision rules on the input features to predict the class labels \cite{dtc}. SVC, in contrast, tries to predict the hyperplane fitting the target variable by maximizing the margin and keeping the error within a threshold \cite{svc}. Therefore, only the support vectors residing in the margin contribute to the decision boundary and determine the error tolerance of the fitted hyperplane. MLP is a feed-forward type NN consisting of one input layer, one or more hidden layers, and one output layer. It is trained by using a backpropagation algorithm which iterates backwards the errors from the output layer to the lower layers, and feed-forwards the weight updates from the input layer to the higher layers \cite{mlp}. Different from MLP, RNN is a NN that utilizes an internal memory component to remember its previous outputs to be used as next inputs which enables it to appropriately model sequential data \cite{rnn}. CNN, on the contrary, is a regularized form of MLP where fully connected relations are replaced with shift invariant and weight shared convolution filters which allows it to better model the spatial and temporal correlations of the data \cite{cnn}.
	
	The detection rate (DR), false alarm rate (FA), and F1 score of  each model for each test system are given as percentages in Table~\ref{tab:results}. Clearly, the LNR-based BDD system falls behind every other model due to non-separable class distributions of $\bm{r}^N$ values, as depicted in Fig.~\ref{fig:class}. It simply predicts each sample as malicious, and this results in  100\% FA rate for each test system. Non-NN based approaches such as DTC and SVM, in contrast, enhance the FA and perform better than BDD by an F1 score range between 67.91\% - 85.97\% due to  their nonlinear modeling capabilities. The NN-based family surpasses the non-NN based models in general, except the MLP where it achieves comparable results with SVC and DTC. The RNN-based detector yields  86.33\%, 83.87\%, and 71.08\% F1 score for IEEE 14, 118, and 300 bus systems, respectively. Only CNN and GNN based detectors reach the 90\% F1 range. Nevertheless, GNN outperforms CNN models by 3.14\%, 4.25\% and 4.41\% in F1 for IEEE test cases with 14, 118, and 300 buses, respectively.
	
	\begin{table}[h]
		\centering
		\newcolumntype{?}[1]{!{\vrule width #1}}
		\setlength{\tabcolsep}{3.0pt}
		\renewcommand{\arraystretch}{1.1}
		{\color{black}
		\caption{Comparison of detector performances (best in bold, worst in italic) in terms of detection rate (DR), false alarm (FA), and F-measure (F1) classification metrics for each IEEE test case system with 14-, 118-, and 300-bus test systems where.}
		\begin{tabular}{c?{1pt}c|c|c?{1pt}c|c|c?{1pt}c|c|c}
			& \multicolumn{3}{c?{1pt}}{\textbf{IEEE 14}} & \multicolumn{3}{c?{1pt}}{\textbf{IEEE 118}} & \multicolumn{3}{c}{\textbf{IEEE 300}} \\ \hline
			\textbf{model} &\textbf{DR}&\textbf{FA}&\textbf{F1}&\textbf{DR}&\textbf{FA}&\textbf{F1}&\textbf{DR}&\textbf{FA}&\textbf{F1} \\ \hline
			\textbf{BDD} & \textbf{100.0} & \textit{100.} & \textit{27.35} &  \textbf{100.0} & \textit{100.} & \textit{26.32} &  \textbf{100.}  & \textit{100.} & \textit{23.26} \\ \hline
			\textbf{DTC} & \textit{68.64} & 29.0 & 67.91 &  75.63 & 22.4 & 77.06 &  74.19 & 9.69 & 78.79 \\ \hline 
			\textbf{SVC} & 75.49 & \textbf{0.13} & 85.97 &  \textit{67.21} & 11.2 & 74.53 &  64.84 & 11.6 & 71.08 \\ \hline
			\textbf{MLP} & 79.42 & 2.95 & 87.17 &  79.65 & 8.86 & 83.89 &  60.83 & 7.51 & 70.54 \\ \hline
			\textbf{RNN} & 78.38 & 3.31 & 86.33 &  74.73 & 3.05 & 83.87 &  64.49 & 9.73 & 71.08 \\ \hline
			\textbf{CNN} & 79.30 & 3.25 & 87.00 &  86.33 & 4.88 & 89.82 &  92.11 & 3.33 & 93.26 \\ \hline
			\textbf{GNN} & 83.97 & 2.43 & \textbf{90.14} &  90.61 & \textbf{1.18} & \textbf{94.07} &  96.51 & \textbf{0.72} & \textbf{97.67} \\ 
		\end{tabular}
		\label{tab:results}
		}
	\end{table}
	
	Our experiments point out that architectural differences in the NN family play a vital role in terms of detection performance.  MLP-based detectors tend to overfit the training data and fail to generalize due to its fully connected relationship between its units. RNNs, in contrast, can not achieve desired results since node values do not form a sequence type of data. Performance of CNN-based models comes after GNN due to their ability to model the temporal and spatial relations of the input data in the Euclidean space, where the locality of the input features can be represented by regular linear grids such as in image or video data. Nevertheless, the inherent graph structure of power grid measurements can not be modeled in the Euclidean space except in trivial cases. As a matter of fact, graph data requires topology-aware models such as GNN to better reflect the adjacency relations of the measurement data.

	\subsection{Impact of Different Weights in the Attack Generation}
\begin{figure}[h!]
		\centering
		\begin{subfigure}{.48\textwidth}
			\centering
3			\includegraphics[width=.99\linewidth]{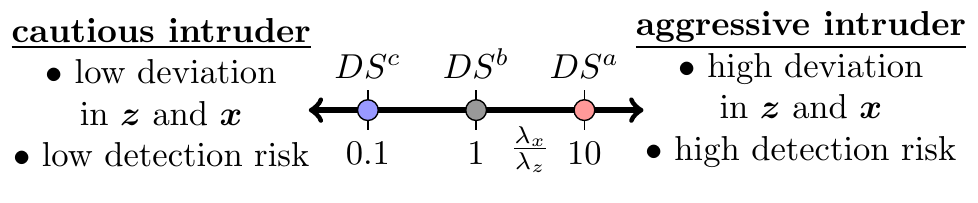}
			\caption{the impact of attack power $\frac{\lambda_{x}}{\lambda_{z}}$ on the generated datasets.}
			\label{fig:tradeoff}
		\end{subfigure}
		\begin{subfigure}{.24\textwidth}
			\centering
			\includegraphics[width=.99\linewidth]{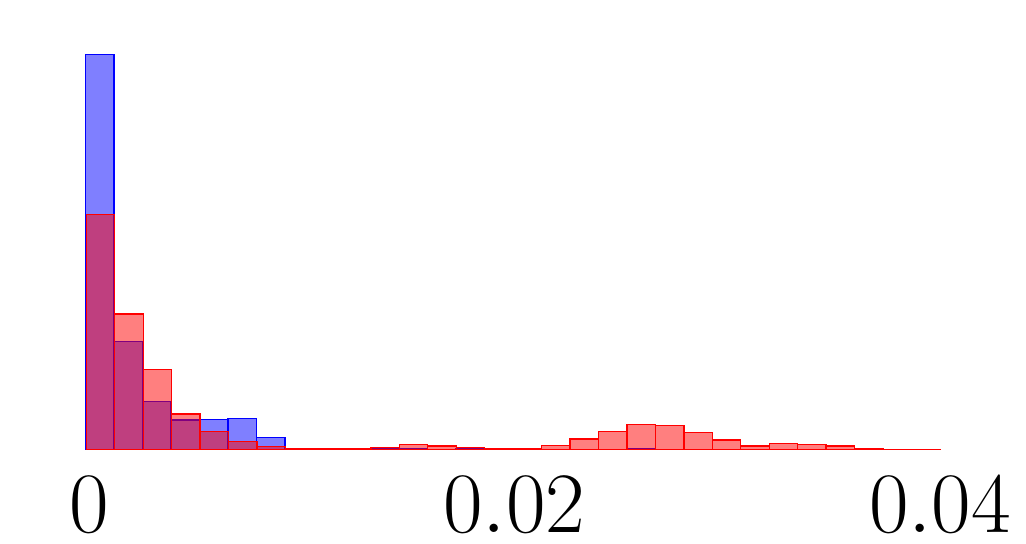}
			\caption{$\max \Delta |V|$ [p.u.].}
			\label{fig:Vm}
		\end{subfigure}
		\begin{subfigure}{.24\textwidth}
			\centering
			\includegraphics[width=.99\linewidth]{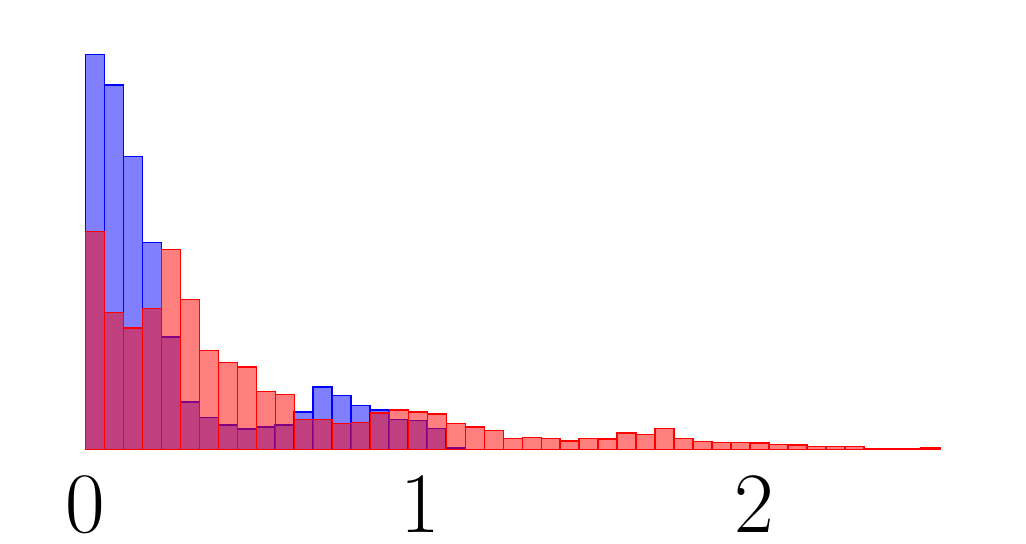}
			\caption{$\max \Delta |\theta|$ [degree].}
			\label{fig:Va}
		\end{subfigure}
		\caption{The impact of attack power on the datasets and distributions of maximum absolute difference of state variables due to attacks in $DS^c$ (blue) and $DS^a$ (red) in terms of voltage magnitude (\ref{fig:Vm}) and angle (\ref{fig:Va}) obtained by Algorithm~\ref{alg:attack}. Note that while max $\Delta |V|$ and $\Delta |\theta|$ values can reach 0.01 p.u. and 1.1 degrees in $DS^c$, they are spread up to 0.04 p.u. and 2.5 degrees in $DS^a$, respectively.}
		\label{fig:attack_dataset}
	\end{figure}

	To assess the impact of different weights in eq. (\ref{eq:attack_opt}) on detection performance of the proposed approach, we generate two extra datasets for each test system having 14-, 118-, and 300-bus test systems.
	In this connection, the Algorithm \ref{alg:attack} is executed two more times for each test system with $\lambda_{z}=10, \ \lambda_{x}=1$ for dataset-c ($DS^c$) and $\lambda_{z}=1, \ \lambda_{x}=10$ for dataset-a $DS^a$ as specified in line 4 of Algorithm \ref{alg:attack} where $DS^c$ and $DS^a$ indicates \textbf{cautious} and \textbf{aggressive} intruder, respectively.
	In $DS^c$, the intruder becomes more cautious and avoids potential detection by decreasing $\frac{\lambda_{x}}{\lambda_{z}}$ ratio which corresponds to the attack power.
	In contrast, the intruder becomes more aggressive and increases the attack power by increasing this ratio at the expense of high detection risk in $DS^a$.
	S/he keeps weighting factors \textbf{balanced} in previously generated dataset-b ($DS^b$) by assigning $\lambda_{z}=1, \ \lambda_{x}=1$. Fig.~\ref{fig:attack_dataset} illustrates the dataset distributions corresponds to $DS^c$ and  $DS^a$ with blue and red colors, respectively.
	It can be seen from Fig.~\ref{fig:attack_dataset} that the ratio of weighting factors $\lambda_{x}$ and $\lambda_{z}$ defined in (\ref{eq:attack_opt}) directly affect the attack power on the state variables.

\begin{figure}[h!]
		\centering
		\newcommand{\wdt}{0.24} 
		\begin{subfigure}{\wdt\textwidth}
			\centering
			\includegraphics[width=.99\linewidth]{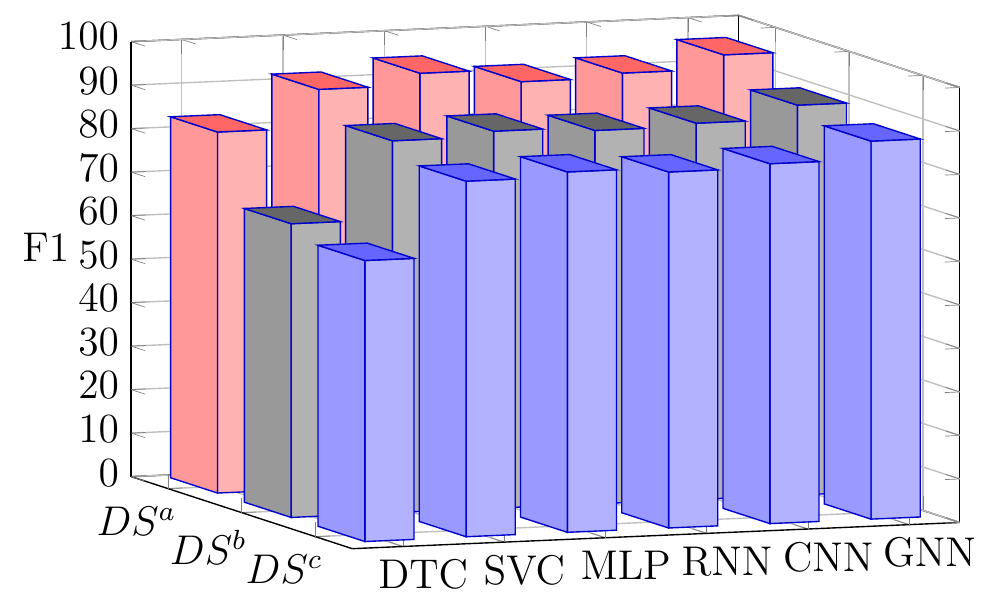}
			\caption{Effect of attack power and detection model for $n=14$}
			\label{fig:effect_014}
		\end{subfigure}
		\begin{subfigure}{\wdt\textwidth}
			\centering
			\includegraphics[width=.99\linewidth]{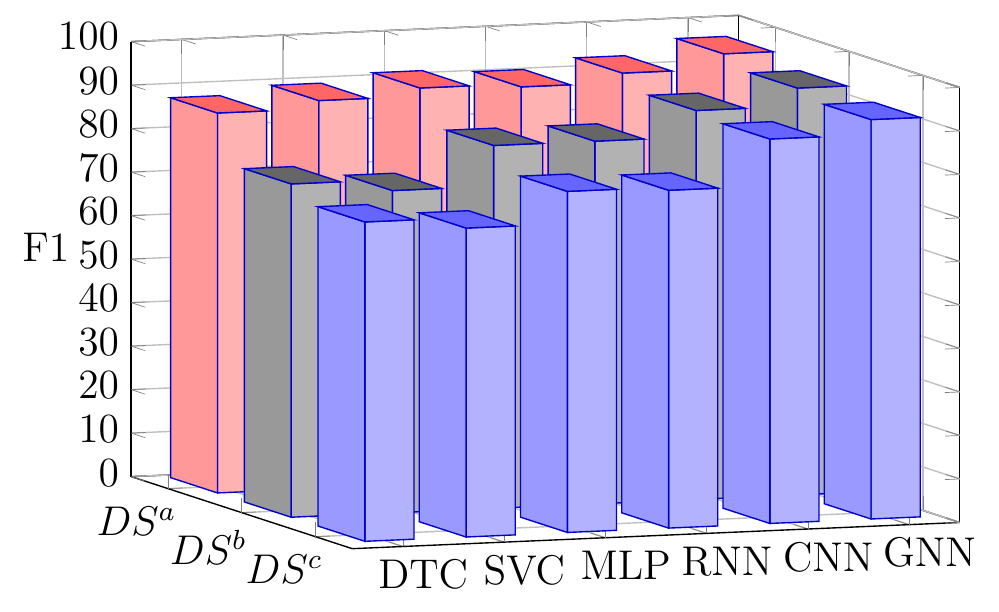}
			\caption{Effect of attack power and detection model for $n=118$}
			\label{fig:effect_118}
		\end{subfigure}
		\begin{subfigure}{\wdt\textwidth}
			\centering
			\includegraphics[width=.99\linewidth]{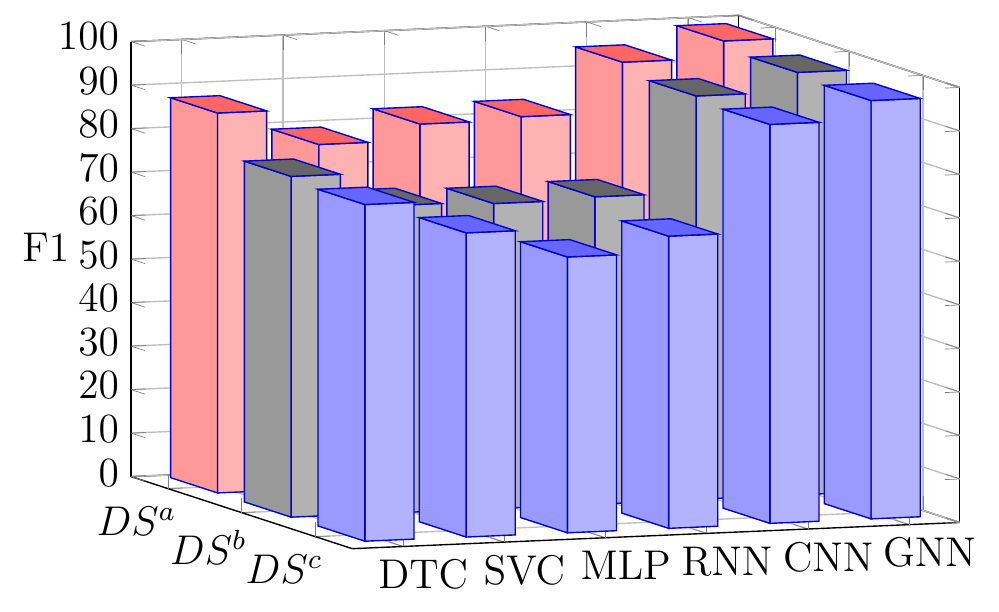}
			\caption{Effect of attack power and detection model for $n=300$}
			\label{fig:effect_300}
		\end{subfigure}
		\begin{subfigure}{\wdt\textwidth}
			\centering
			\includegraphics[width=.99\linewidth]{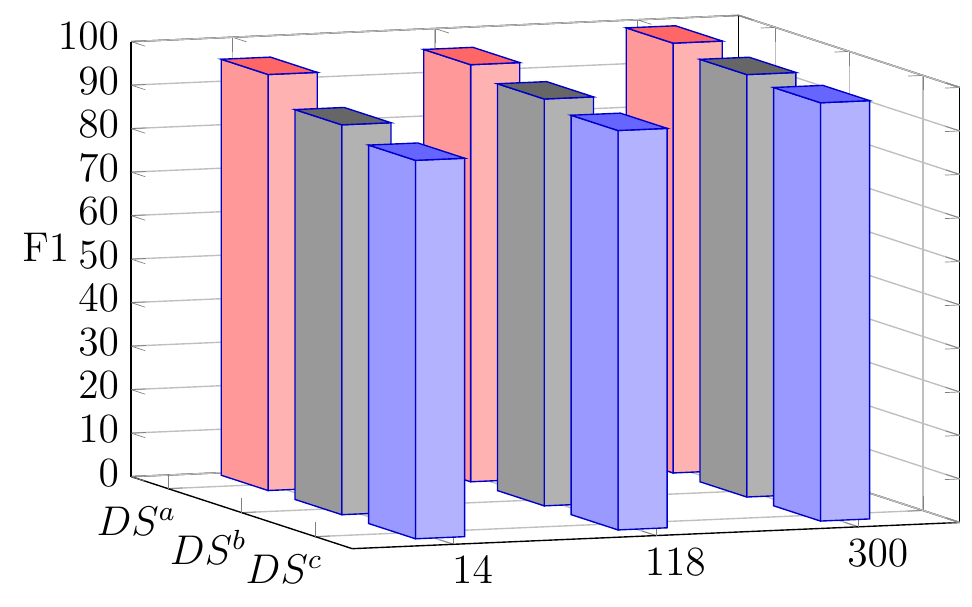}
			\caption{Effect of attack power and number of buses for GNN model}
			\label{fig:effect_gnn}
		\end{subfigure}
		\caption{The impact of attack power, detector model, and grid size on the F1 score of implemented detectors. For each model and test system, performance of the detectors increase from `cautious' ($DS^c$) to `aggressive' ($DS^a$) attacks in Figs. \ref{fig:effect_014}, \ref{fig:effect_118}, and \ref{fig:effect_300}, for IEEE 14-, 118-, and 300-bus test systems, respectively. In Fig. \ref{fig:effect_gnn}, effect of attack power and grid side is visualized for the proposed GNN based detector. Performance of the detector for the proposed model increases with larger systems and more aggressive attacks.}
		\label{fig:effect}
	\end{figure}

 	After obtaining $DS^c$ and $DS^a$ for each test system, we applied our detectors on $DS^c$ and $DS^a$ to compare model performances similar to the previous comparisons conducted on $DS^b$.
 	Namely, we split and scale the datasets and train detector models on the training split, optimize the parameters on the validation split and evaluate the final results on the test split.
 	Fig.~\ref{fig:effect} summarizes the classification results in terms of F1 ratios.
 	As expected, detection performances increase from $DS^c$ to $DS^a$ for each model and test system due to comparably more separable class distributions between honest and malicious samples.
	However, the proposed GNN outperforms the best available solutions in the literature for 14-, 118-, and 300-bus test systems by 4.31\%, 3.46\%, 4.48\% for the `cautious', 3.14\%, 4.25\%, 4.41\% for the `balanced', and 3.19\%, 3.43\%, 3.88\% for the `aggressive' intrusion, respectively.
 	In addition, it is observed from our experiments that GNN performs better compared to other models in larger cases because when the number of nodes increases, the spatial correlation between adjacent measurements becomes more dominant compared to the global correlations between all measurements. Since GNN is specifically designed to exploit this spatial information of the data, it performs a better job for larger cases. In other words, the denser topology translates into more spatial correlation which improves GNN's accuracy.
 
	\subsection{Tuning Model Hyperparameters}	
	Traditional hyper-parameter tuning algorithms such as random search and grid search try to find the optimal parameters by randomly or sequentially sampling the parameters from the hyper-parameter space of the model and ignoring the results of  previous trials. For example, even though any combination of a parameter set for a specific value of a parameter would fail to perform well regardless of the other parameters, these techniques might continue to run its trials with these specific values. Thereupon, selecting the optimal hyper parameters using random or grid search could be highly stagnant particularly for large hyper-parameter spaces. On the contrary, the Bayesian optimization technique considers performances of the past trials to better explore the parameter space. By focusing on more `promising' regions of the parameters space in the light of its past experiences, it tries to select parameter combinations which give better validation performance. As a consequence of this informed navigation and sampling, it reduces the search time and offers a better set of parameters which can lead to better model performance \cite{machine_learning}.

	All model hyper-parameters are tuned using Bayesian optimization techniques, Sklearn \cite{sklearn} and  Keras-tuner \cite{kerastuner} Python libraries. Whereas the model fitting is performed on the training split of  data, evaluation and optimal parameter selection are carried out on the validation split. After choosing the best hyper-parameters in 200 trials for each model, performances of the models having optimal parameters  are assessed on the test splits, and results are saved. Please refer to Table~\ref{tab:hpo} for the hyper-parameters, their space, and optimal values for each model and test system.

	\begin{table}[h] 
		\centering
		\setlength{\tabcolsep}{2pt}
		\newcolumntype{?}[1]{!{\vrule width #1}}
		{\color{black}
		\caption{Optimized model hyper-parameters.}
		\begin{tabular}{c?{1pt}c?{1pt}c?{1pt}c?{1pt}c?{1pt}c}
			\textbf{model}  & \textbf{param} & \textbf{space} & \textbf{IEEE-14} & \textbf{IEEE-118} & \textbf{IEEE-300} \\
			\specialrule{1pt}{1pt}{1pt}
			
			\textbf{BDD}
			& threshold		&  \{0.01, 0.02, \dots, 5.0\}	& 1.05    & 2.37    & 2.62    \\ 
			\specialrule{1pt}{1pt}{1pt}

			\multirow{4}{*}{\textbf{DTC}}  
			& criterion     & \{gini, entropy\}   			& entropy & gini    & gini    \\ \cline{2-6}
			& depth         & \{8, 9,\dots, 64\}   			& 64      & 64      & 64      \\ \cline{2-6}
			& features      & \{0.1, 0.2, \dots, 0.9\} 		& 0.3     & 0.4     & 0.5     \\ \cline{2-6}
			& min. leaf     & \{1, 2, \dots, 8\} 			& 4       & 1       & 2       \\ 
			\specialrule{1pt}{1pt}{1pt}
			
			\multirow{4}{*}{\textbf{SVC}}
			& C             & $10^{\{-6, -5, \dots, 2\}}$	& $10^2$  & $10^2$  & $10^{1}$\\ \cline{2-6} 
			& degree        & \{1, 2, \dots, 5\}    		& 2       & 2       & -       \\ \cline{2-6} 
			& gamma         & $10^{\{-6, -5, \dots, 2\}}$   &$10^{-1}$&$10^{-3}$&$10^{-3}$\\ \cline{2-6} 
			& kernel        & \{linear, poly, rbf\}         & poly    & poly     & rbf    \\
			\specialrule{1pt}{1pt}{1pt}
			
			\multirow{4}{*}{\textbf{MLP}}  
			& layers        & \{1, 2, 3, 4\}        		& 4       & 3       & 3       \\ \cline{2-6} 
			& units         & \{8, 16, 32, 64\}  		    & 16      & 16      & 64      \\ \cline{2-6}
			& activation    & \{relu, elu, tanh\} 			& elu     & elu     & elu     \\ \cline{2-6}
			& optimizer     & \{adam, sgd, rmsprop\}        & rmsprop & adam    & rmsprop \\
			\specialrule{1pt}{1pt}{1pt} 
			
			\multirow{4}{*}{\textbf{RNN}}  
			& layers        & \{1, 2, 3, 4\}        		& 3       & 4       & 4       \\ \cline{2-6} 
			& units         & \{8, 16, 32, 64\}  	     	& 16      & 32      & 16      \\ \cline{2-6}
			& activation    & \{relu, elu, tanh\} 			& relu    & relu    & relu    \\ \cline{2-6}
			& optimizer     & \{adam, sgd, rmsprop\}        & adam    & adam    & rmsprop \\
			\specialrule{1pt}{1pt}{1pt} 
								
			\multirow{4}{*}{\textbf{CNN}}  
			& layers        & \{1, 2, 3, 4\}        		& 3       & 2       & 3       \\ \cline{2-6} 
			& units         & \{8, 16, 32, 64\}     		& 16      & 16      & 32      \\ \cline{2-6}
			& K             & \{2, 3, 4, 5\}           		& 5       & 5       & 5       \\ \cline{2-6}
			& activation    & \{relu, elu, tanh\} 			& relu    & relu    & relu    \\ \cline{2-6}
			& optimizer     & \{adam, sgd, rmsprop\}        & rmsprop & adam    & adam    \\
			\specialrule{1pt}{1pt}{1pt} 					
						
			\multirow{4}{*}{\textbf{GNN}}  
			& layers        & \{1, 2, 3, 4\}        		& 3       & 3       & 4       \\ \cline{2-6} 
			& units         & \{8, 16, 32, 64\}      		& 32      & 16      & 32      \\ \cline{2-6}
			& K             & \{2, 3, 4, 5\}           		& 3       & 3       & 2       \\ \cline{2-6}
			& activation    & \{relu, elu, tanh\} 			& relu    & relu    & relu    \\ \cline{2-6}
			& optimizer     & \{adam, sgd, rmsprop\}        & adam    & adam    & adam    \\
		\end{tabular}
		\label{tab:hpo}
		}
	\end{table}

	\section{Conclusion}
	In this paper, we addressed the detection of stealth FDIA in modern AC power grids. To that end, we first developed a generic, locally applied, and stealth FDIA generation technique by solving a nonlinear  non-convex optimization problem using SGD algorithm and made available the labeled data to the research  community. Second, we proposed a scalable and real-time detection mechanism for FDIAs by fusing the underlying graph topology of the power grid and spatially correlated measurement data in GNN layers. Finally, we tested our algorithms on standard test beds such as IEEE 14-, 118-, and 300-bus systems and demonstrated that the proposed GNN detector surpasses the currently available methods in literature by 3.14\%, 4.25\% and 4.41\% in F1 score, respectively.	
	
	\bibliographystyle{IEEEtran}
	\bibliography{main}
	
	\begin{IEEEbiography}[{\includegraphics[width=1in,height=1.25in,clip,keepaspectratio]{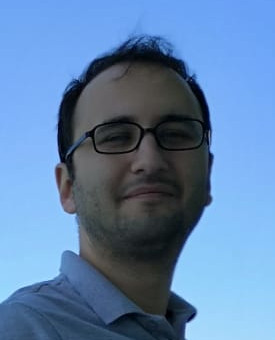}}]%
	{Osman Boyaci} (Student Member, IEEE) received the B.Sc. (Hons.) degree in electronics engineering in 2013 and in computer engineering (double major) in 2013 (Hons.) from Istanbul Technical University, Istanbul, Turkey. He received the M.Sc. degree in computer engineering at the same university in 2017. Currently, he is a Ph.D. candidate at Texas A\&M University working on graph neural network based cybersecurity in smart grids.
	
	His research interests include machine learning, artificial intelligence, and cybersecurity.
	\end{IEEEbiography}
	
	\vspace{-37pt}
	
	\begin{IEEEbiography}[{\includegraphics[width=1in,height=1.25in,clip,keepaspectratio]{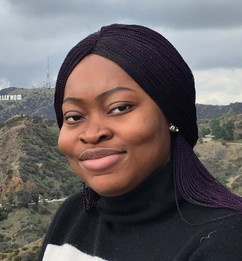}}]%
	{Amarachi Umunnakwe} (Student Member, IEEE) currently a Ph.D. candidate in Electrical and Computer Engineering at Texas A\&M University, College Station. She received her B.S. degree in Electronic Engineering from the University of Nigeria, Nsukka and her M.S. degree in Electrical and Computer Engineering from the University of Utah. Her research interests include cyber-physical resilience, situational awareness, and security of electric power systems using intelligent techniques.
	\end{IEEEbiography}
	
	\vspace{-37pt}
	
	\begin{IEEEbiography}[{\includegraphics[width=1in,height=1.25in,clip,keepaspectratio]{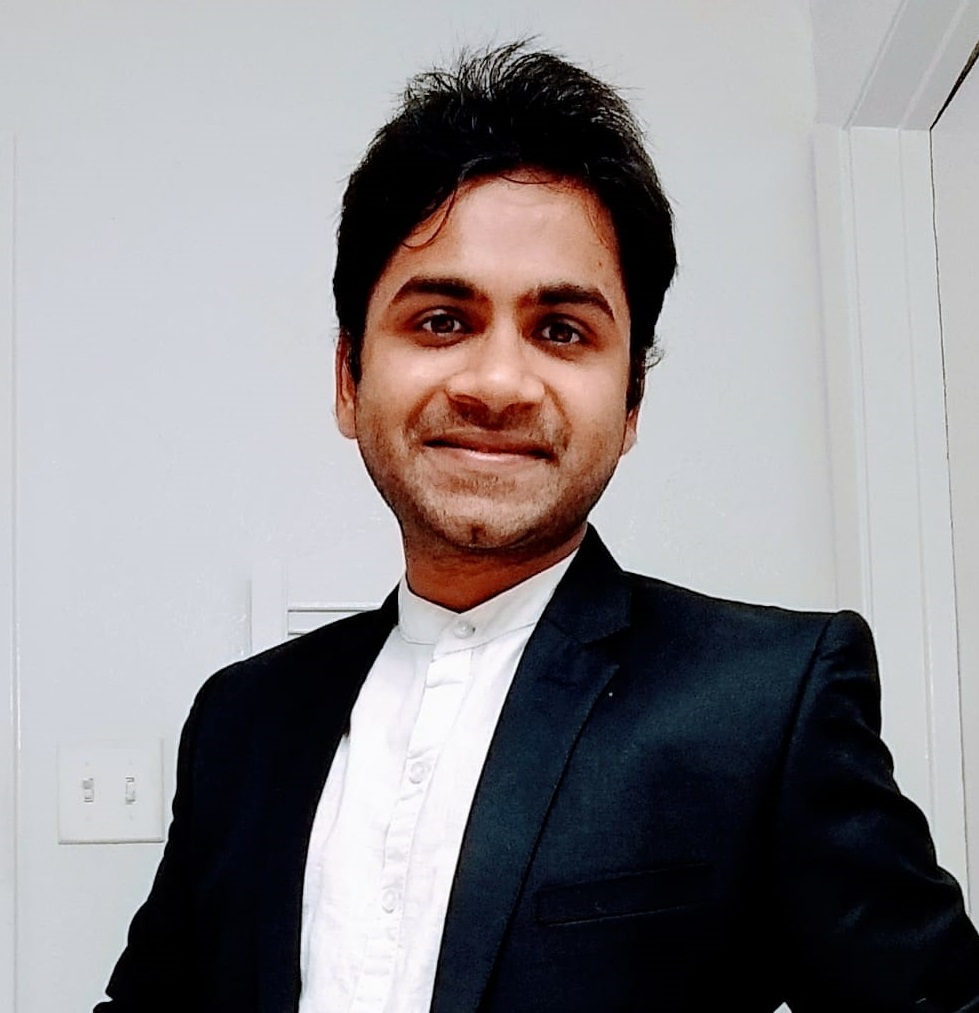}}]%
	{Abhijeet Sahu} (Student Member, IEEE) received his B.S. degree in Electronics and Communications from National Institute of Technology, Rourkela, India, in 2011 and his M.S. degree in Electrical and Computer Engineering from Texas A\&M University, TX, USA in 2018.
	Currently, he is a Ph.D. candidate at Texas A\&M University working on Cyber-Physical Resilient Energy Systems.
	
	His research interests include network security, cyber-physical modeling for intrusion detection and response, and Artificial Intelligence for cyber-physical security in power systems. 
	\end{IEEEbiography}
	
	\vspace{-37pt}
	
	\begin{IEEEbiography}[{\includegraphics[width=1in,height=1.25in,clip,keepaspectratio]{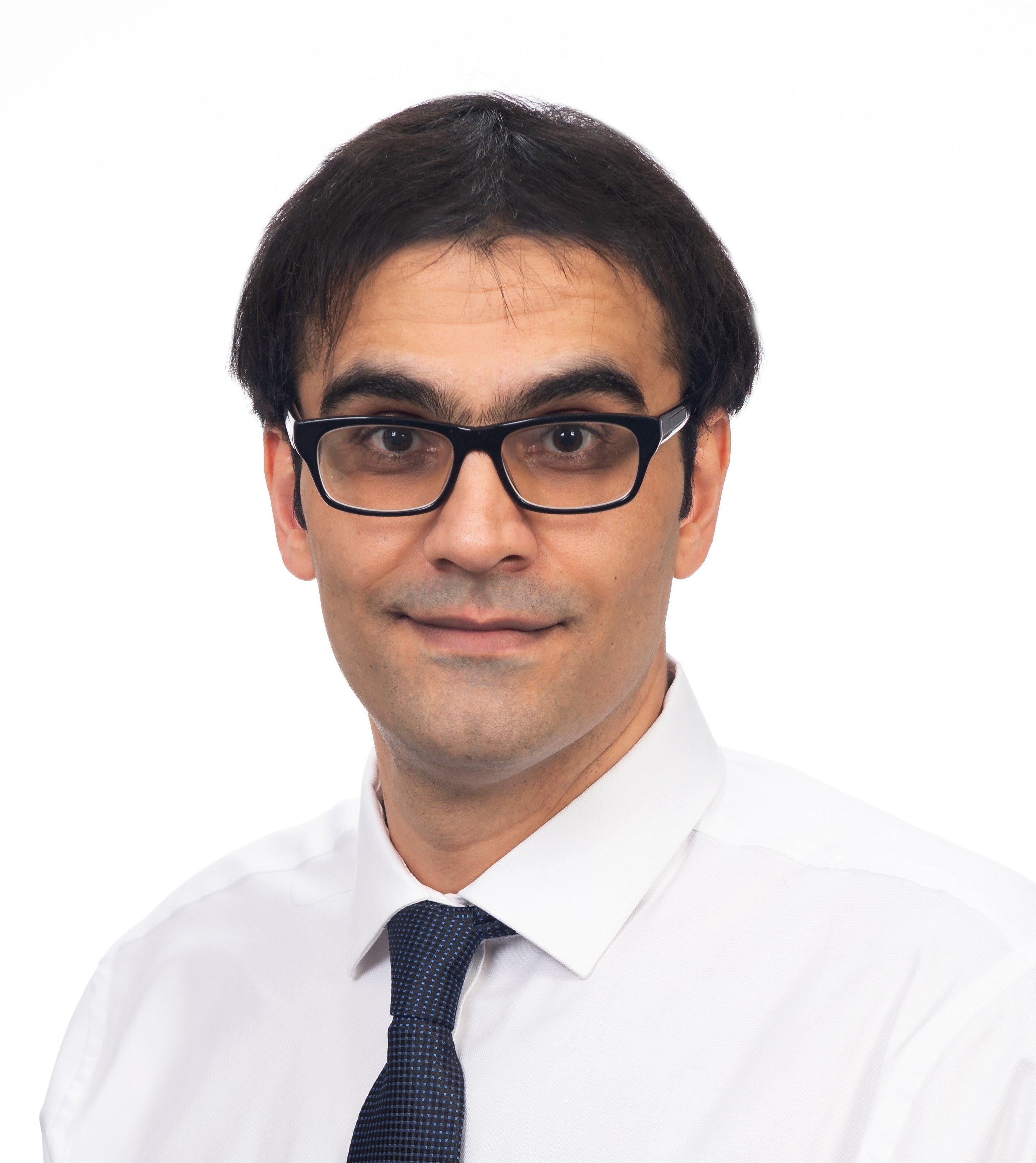}}]%
	{Mohammad Rasoul Narimani}
	(S'14-M'20) is an Assistant Professor in the College of Engineering at Arkansas State University. Before joining Arkansas State University, he was a postdoc at Texas A\&M University, College Station. He received the B.S. and M.S. degrees in electrical engineering from the Razi University and Shiraz University of Technology, respectively. He received the Ph.D. in electrical engineering from Missouri University of Science \& Technology. His research interests are in the application of optimization techniques to electric power systems.
	\end{IEEEbiography}
	
	\vspace{-37pt}
	
	\begin{IEEEbiography}[{\includegraphics[width=1in,height=1.25in,clip,keepaspectratio]{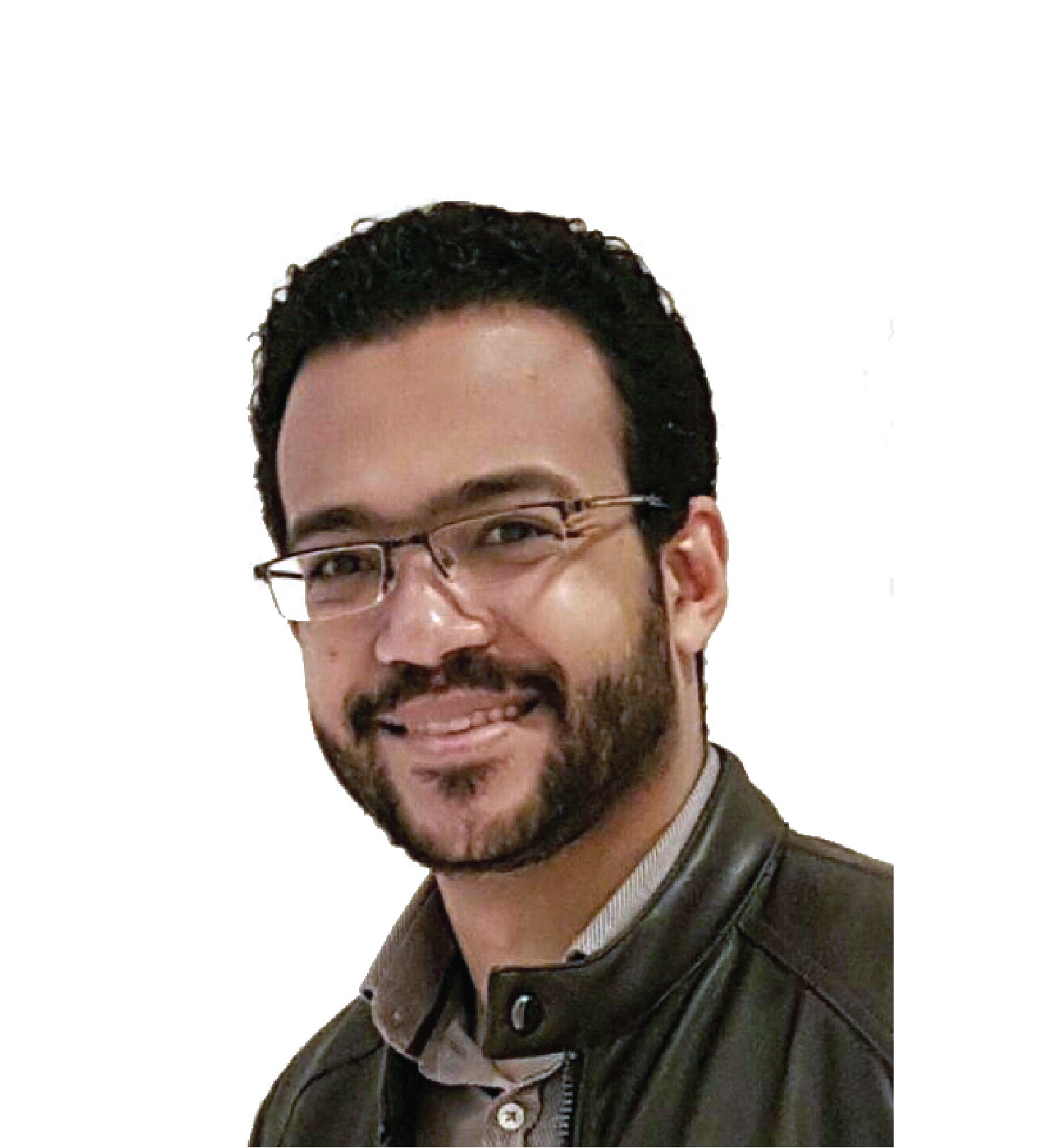}}]%
	{Muhammad Ismail} (S'10-M'13-SM'17) received the B.Sc. (Hons.) and M.Sc. degrees in Electrical Engineering (Electronics and Communications) from Ain Shams University, Cairo, Egypt, in 2007 and 2009, respectively, and the Ph.D. degree in Electrical and Computer Engineering from the University of Waterloo, Waterloo, ON, Canada, in 2013. He is currently an Assistant Professor with the Department of Computer Science, Tennessee Tech. University, Cookeville, TN, USA. He was a co-recipient of the Best Paper Awards in the IEEE ICC 2014, the IEEE Globecom 2014, the SGRE 2015, the Green 2016, the Best Conference Paper Award from the IEEE TCGCN at the IEEE ICC 2019, and IEEE IS 2020.  
	\end{IEEEbiography}
	
	\vspace{-25pt}
	
	\begin{IEEEbiography}[{\includegraphics[width=1in,height=1.25in,clip,keepaspectratio]{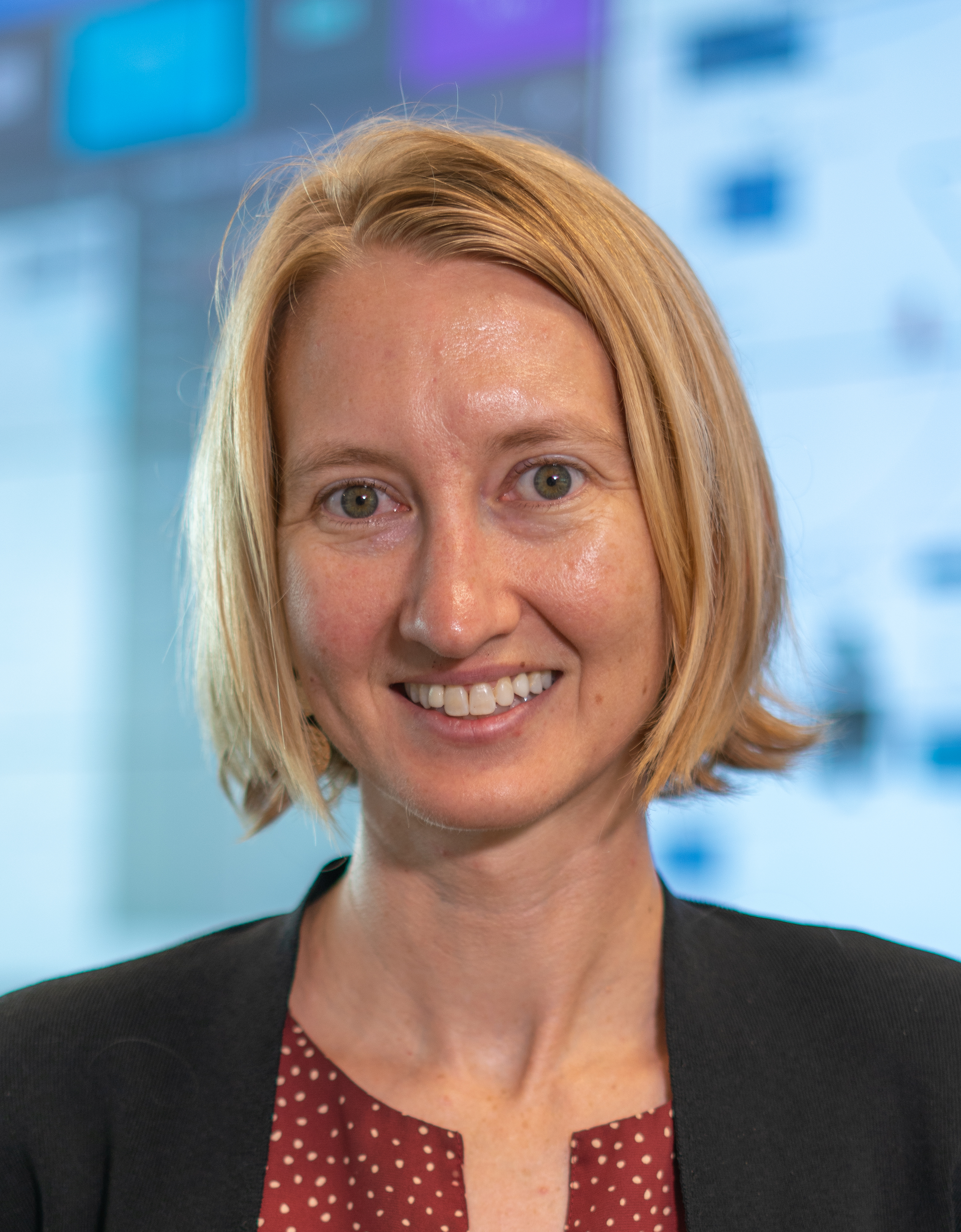}}]%
	{Katherine R. Davis} (S'05-M'12-SM'18) received the B.S. degree from the University of Texas atAustin, Austin, TX, USA, in 2007, and the M.S. and Ph.D. degrees from the University of Illinois at Urbana-Champaign, Champaign, IL, USA, in 2009 and 2011, respectively, all in electrical engineering. She is currently anAssistant Professor of Electrical and Computer Engineering at TAMU.
	\end{IEEEbiography}
	
	\vspace{-20pt}
	
	\begin{IEEEbiography}[{\includegraphics[width=1in,height=1.25in,clip,keepaspectratio]{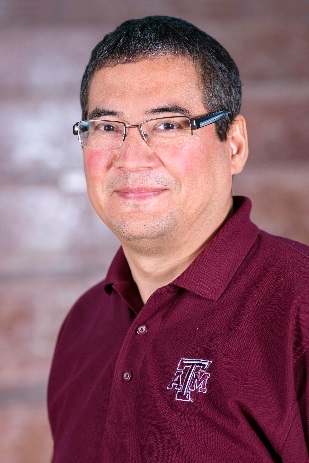}}]%
	{Dr. Erchin Serpedin} is a professor in the Electrical and Computer Engineering Department at Texas A\&M University in College Station. Dr. Serpedin is the author of four research monographs, one textbook, 17 book chapters, 170 journal papers, and 270 conference papers. His current research interests include signal processing, machine learning, artificial intelligence, cyber security, smart grids, and wireless communications. He served as an associate editor for more than 12 journals, including journals such as the IEEE Transactions on Information Theory, IEEE Transactions on Signal Processing, IEEE Transactions on Communications, IEEE Signal Processing Letters, IEEE Communications Letters, IEEE Transactions on Wireless Communications, IEEE Signal Processing Magazine, and Signal Processing (Elsevier), and as a Technical Chair for six major conferences. He is an IEEE Fellow. 
	\end{IEEEbiography}
	
\end{document}